\documentclass[secnumarabic,amssymb,nobibnotes,aps,prb,preprint,letterpaper,preprintnumbers,amsmath,superscriptaddress,endfloats*]{revtex4-2}
\usepackage{graphicx}
\usepackage{dcolumn}
\usepackage{bm}
\usepackage{amsmath,amssymb,bm}
\usepackage{graphicx}
\usepackage{xspace}
\usepackage{latexsym}
\usepackage{color}
\usepackage{lineno}
\usepackage{xr}
\usepackage{siunitx}
\usepackage[normalem]{ulem}
\usepackage{soul}
\usepackage [latin1]{inputenc}
\usepackage[colorlinks = true,linkcolor = blue,urlcolor  = red,citecolor = red,anchorcolor = blue]{hyperref}
\usepackage{placeins}
\usepackage{orcidlink}
\usepackage{glossaries}
\usepackage{cleveref}
\usepackage{gensymb}

\newacronym{INS}{INS}{inelastic neutron scattering}
\newacronym{RIXS}{RIXS}{resonant inelastic x-ray scattering}
\newacronym{REXS}{REXS}{resonant elastic x-ray scattering}
\newacronym{XRD}{XRD}{synchrotron x-ray diffraction}
\newacronym{QSL}{QSL}{quantum spin liquid}
\newacronym{DSF}{DSF}{dynamical structure factor}
\newacronym{MC}{MC}{magnetic continuum}
\newacronym{XAS}{XAS}{x-ray absorption spectroscopy}
\newacronym{XMCD}{XMCD}{x-ray magnetic circular dichroism}
\newacronym{ARPES}{ARPES}{angular resolved photoemission spectroscopy}
\newacronym{STM}{STM}{scanning tunneling microscopy}

\newacronym{SEM}{SEM}{scanning electron microscopy}
\newacronym{OSMT}{OSMT}{orbital selective Mott transition}
\newacronym{TEY}{TEY}{total electron yield}
\newacronym{LH}{LH}{linear horizontal}
\newacronym{LV}{LV}{linear vertical}
\newacronym{CDW}{CDW}{charge density wave}
\newacronym{CO}{CO}{charge order}
\newacronym{2D}{2D}{two-dimensional}
\newacronym{3D}{3D}{three-dimensional}
\newacronym{BZ}{BZ}{brillouin zone}

\newcommand{\FGT} {Fe$_{4.75}$GeTe$_2$\xspace}
\newcommand{\FfGT} {Fe$_{5}$GeTe$_2$\xspace}
\newcommand{\FfxGT} {Fe$_{5-x}$GeTe$_2$\xspace}

\newcommand{\FethGT} {Fe$_{3}$GeTe$_2$\xspace}

\newcommand{\FetsGT} {Fe$_{2.72}$GeTe$_2$\xspace}

\newcommand{\FeGeTe} {Fe-Ge-Te\xspace}

\newcommand{\NSLSII} {National Synchrotron Light Source II, Brookhaven National Laboratory, Upton, New York 11973, USA.\xspace}
\newcommand{\APS}{Advanced Photon Source, Argonne National Laboratory, Lemont, IL, USA.\xspace}
\newcommand{\ColumbiaUni}{Department of Physics, Columbia University, New York, New York 10027, USA.\xspace}

\newcommand{\vb}[1]{\textcolor{black}{#1}}

\begin{document}



\title{Magnetic excitations and absence of charge order in the van der Waals ferromagnet \FGT}

\author{V. K. Bhartiya}
\email{vbhartiya1@bnl.gov}
\affiliation{\NSLSII}

\author{T. Kim}%
\affiliation{\NSLSII}
\author{J. Li}%
\affiliation{\NSLSII}

\author{T. P. Darlington}%
\affiliation{\ColumbiaUni}
\author{D. J. Rizzo}%
\affiliation{\ColumbiaUni}

\author{Y. Gu}%
\affiliation{\NSLSII}
\author{S. Fan}%
\affiliation{\NSLSII}

\author{C. Nelson}
\affiliation{\NSLSII}
\author{J. W. Freeland}
\affiliation{\APS}

\author{X. Xu}%
\affiliation{Department of Physics, University of Washington, Seatle, Washington 98195, USA.}
\author{D. N. Basov}%
\affiliation{\ColumbiaUni}
\author{J. Pelliciari}%
\affiliation{\NSLSII}
\author{A. F. May}%
\affiliation{Materials Science and Technology Division, Oak Ridge National Laboratory, Oak Ridge, Tennessee 37831, USA.}

\author{C. Mazzoli}
\affiliation{\NSLSII}
\author{V. Bisogni}%
\email{bisogni@bnl.gov}
\affiliation{\NSLSII}

\date{\today}
\maketitle

\textbf{Understanding the ground state of van der Waals (vdW) magnets is crucial for designing devices leveraging these platforms. Here, we investigate the magnetic excitations and charge order in \FGT, a vdW ferromagnet with $\approx$ 315 K Curie temperature. Using Fe $L_3 - $edge resonant inelastic x-ray scattering, we observe a dual nature of magnetic excitations, comprising a coherent magnon and a broad non-dispersive continuum extending up to 150 meV, 50$\%$ higher than in \FetsGT. The 
continuum intensity is sinusoidally modulated along the stacking direction $L$, with a period matching the inter-slab distance. Our results indicate that while the dual character of the magnetic excitations is generic to Fe-Ge-Te vdW magnets, \FGT exhibits a longer out-of-plane magnetic correlation length, suggesting enhanced 3D magnetic character. Furthermore, resonant x-ray diffraction reveals that previously reported $
\pm$(1/3, 1/3, $L$) peaks originate from crystal structure rather than from charge order.}
\vspace{2cm}

\section{\label{sec:intro}Introduction}
The discovery of ferromagnetism in two-dimensional van der Waals (vdW) ferromagnets have opened exciting frontiers in quantum matter research \cite{Burch2018MagnetismMaterials,Gibertini2019MagneticHeterostructures,Wang2021TheMaterials}. In this context, the 
ternary \FeGeTe vdW materials, such as \FethGT and \FfGT , hosting high-temperature long-range ferromagnetic order down to the \gls*{2D} structural limit is a unique platform where miniaturized spintronics devices have already been proposed \cite{May2019FerromagnetismArticle,Zhang2020ItinerantTemperature,Chen2023Above-Room-Temperature2,Wang2018TunnelingHeterostructures,Georgopoulou-Kotsaki2023Significant,Li2018Patterning,Alghamdi2019Highly,Yang2020Highly,Ding2020Observtion,Wu2020Neel,Deng2018Gate-tunableFe3GeTe2}. 
In addition, strong electronic correlations are reported to drive these systems towards exotic ground states, that is, the heavy Fermions/ Kondo physics~\cite{Yun2018Emergence,Bao2022Neutron2}, \gls*{CO}~\cite{Wu2021DirectGeTe_2}, flat bands \cite{Wu2023ReversibleFerromagnet}, optical anomalous hall effect (AHE) \cite{Kato_npj2022}, and \gls*{OSMT}~\cite{Bai2022AntiferromagnetictextGeTe_2}. 

\FethGT ($ 150 < T_{\rm{C}} < 230$ K) is the most thoroughly investigated member of the ternary \FeGeTe vdW magnets. One of the intriguing findings is the coexistence of low energy coherent magnetic excitations (magnons) and a dispersionless non-Stoner type continuum at higher energies \cite{Bai2022AntiferromagnetictextGeTe_2,Bao2022Neutron2,Xu2020}. This dual character of magnetic excitations is beyond the framework of the linear spin wave theory, invoking the  \gls*{OSMT} scenario -- atypical to 3$d$ transition metals  -- to explain the experimental observations \cite{Bai2022AntiferromagnetictextGeTe_2}. Are these dual magnetic excitations, a generic feature of the ternary \FeGeTe  vdW materials? If so, how \vb{do} these excitations evolve with $T_{\rm C}$? 

\FGT with the highest $T_{\rm{C}} \approx$ 315 K among \FeGeTe  2D-vdW ferromagnetic materials is an ideal member to investigate these questions. Moreover, it hosts the long-range ferromagnetic order remarkably close to room temperature $\approx$ 270 K in the \gls*{2D} structural limit \cite{May2019FerromagnetismArticle,Zhang2020ItinerantTemperature,Chen2023Above-Room-Temperature2,Georgopoulou-Kotsaki2023Significant}. 
Deciphering the mechanism behind these  high temperature \gls*{2D} vdW ferromagnets is essential to engineer novel heterostructures from   \gls*{2D} layers.
However, despite enormous progress in understanding transport, magnetic and structural properties of \FfxGT \cite{May2019FerromagnetismArticle,Zhang2020ItinerantTemperature,Chen2023Above-Room-Temperature2}, to date, the character of the magnetic excitations, their energy scale, and dimensionality remain unexplored in this material. The \gls*{INS} is a standard technique to address these issues \cite{Bettler2019MagneticPbVO,Bhartiya2021InelasticBaVO} but the small size of the \FGT single crystals poses a challenge. Furthermore, a recent \gls*{ARPES} and \gls*{STM} study reports a charge order ~\cite{Wu2021DirectGeTe_2}, that is shown to compete with the magnetic ordering. However, a bulk-sensitive signature of the \gls*{CO}, its electronic character and propagation vector are unexplored so far. 

Here we present an investigation of the magnetic excitations and charge order in \FGT. To elucidate the character, energy scale, and dimensionality of the magnetic excitations, we employed high-resolution Fe $L_3 - $edge \gls*{RIXS} on a mm sized single crystal \FGT. \gls*{RIXS} complements INS to probe collective spin excitations and, in addition, can measure small sample volumes \cite{Pelliciari2021TuningConfinement} easily accessing the high-energy spectral range where the thermal neutron flux is heavily reduced. Our findings suggests that \FGT hosts dual character of magnetic excitations, similarly to \FetsGT, and  despite being a \gls*{2D} system structurally, it displays a significant three-dimensional (3D) interaction between \FeGeTe slabs that are $\sim$10 \AA \, apart. To investigate the bulk and electronic character of the proposed \gls*{CO}, we used \gls*{XRD} and Fe $K - $edge \gls*{REXS} ~\cite{Abbamonte2004CrystallizationSr14Cu24O41,Chen2016Remarkable4,Shen2021ChargeNickelates}. Our results demonstrate that the observed $\pm$(1/3, 1/3, $L$) peaks are of structural origin, suggesting a doubling of structural unit cell along the $c$-axis.  Overall, our work advances the current understanding of the magnetic properties of \FGT and shed light onto the whole family of \FeGeTe vdW magnets.

\begin{figure*}
\includegraphics[width=0.85\textwidth]{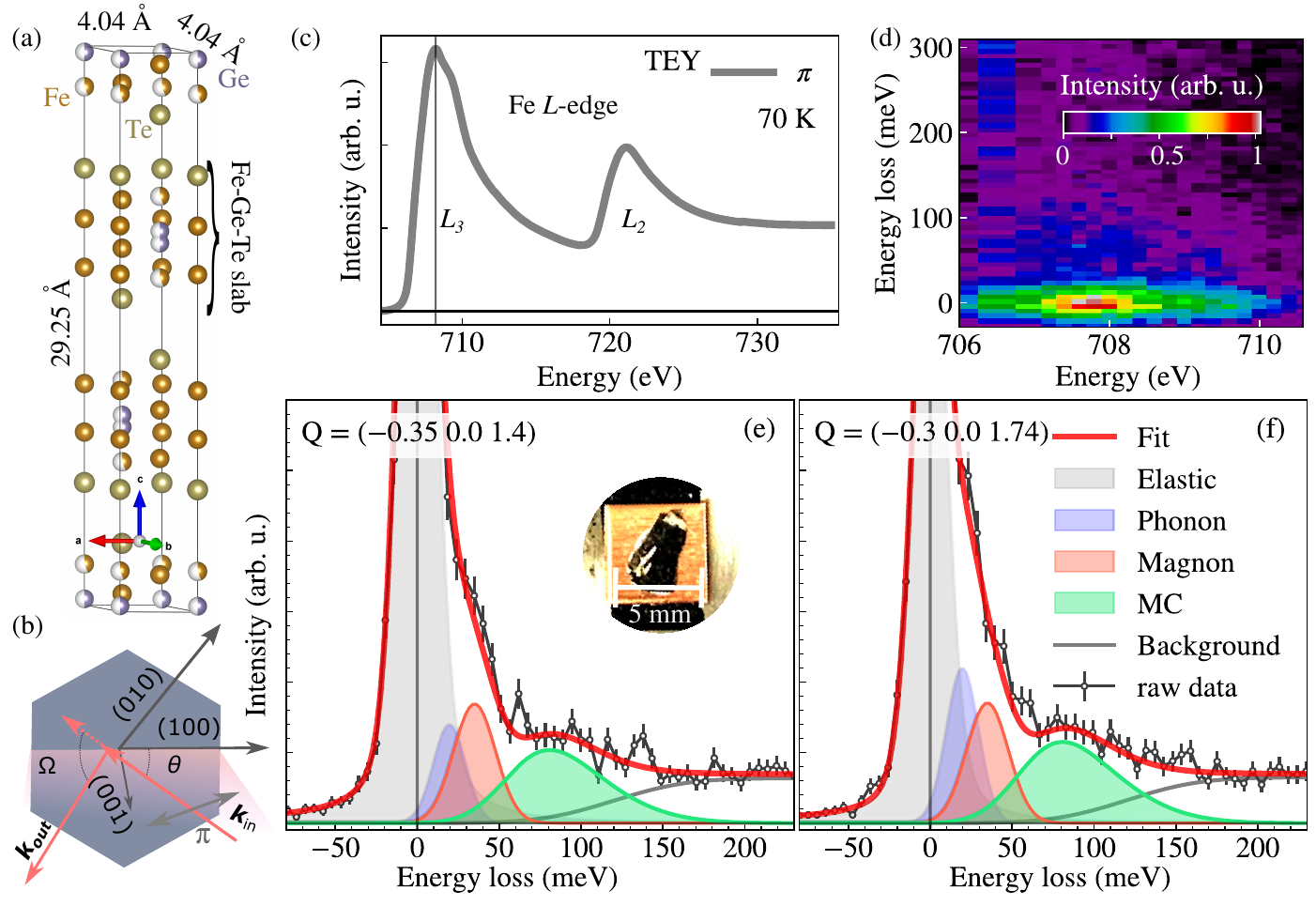}
\caption{Magnetic excitations in the Fe $L_3 - $edge \gls*{RIXS} spectra of \FGT. (a) \FGT crystal structure drawn using VESTA \cite{VESTA}. A single unit cell contains three \FeGeTe slabs. (b) Schematic of \gls*{RIXS} experimental geometry. In the hexagonal structure representation, two orthogonal wave vectors -- (100) and (001) -- form the scattering plane highlighted in pink. $\theta$ is an angle between the incident x-ray and the (100) axis while $\Omega$ is the scattering angle between incoming and outgoing x-ray. The double headed arrow represents the $\pi$ polarization of the incident light. (c) Fe $L_{3, 2} - $edge XAS measured at $\theta = 15^\circ$ in total electron yield (TEY) mode at 70 K. A solid vertical line represents the Fe $L_3 - $edge maximum, that corresponds to the incident photon energy (707.9 eV) used for the \gls*{RIXS} spectra. (d) RIXS intensity map measured as a function of incident photon around the Fe $L_3 - $edge, and zoomed in to cover excitations below 300 meV. For this dataset, $\theta = 20^\circ$ and $\Omega = 95^\circ$ corresponding to $\bm{Q}$ = ($-0.16$ 0 2.17).  (e-f). Two representative \gls*{RIXS} scans (black dotted line) at ($-0.35$ 0 1.4) and ($-0.3$ 0 1.74). The low-energy excitations are fitted with an elastic peak (gray), a phonon (blue), a magnon (red), a broad continuum, and a background (gray line) stretching to higher energies. The red solid line represents the fit sum. The inset in (e) shows the measured \FfGT single crystal. All RIXS spectra displayed in this work were measured at $T$ = 70 K, using $\pi$ polarization. The error bars of the \gls*{RIXS} spectra are defined assuming a Poisson distribution of the single-photon counted events.}\label{FIG:fig1}
\end{figure*}

\begin{figure*}
  \includegraphics[width=\textwidth]{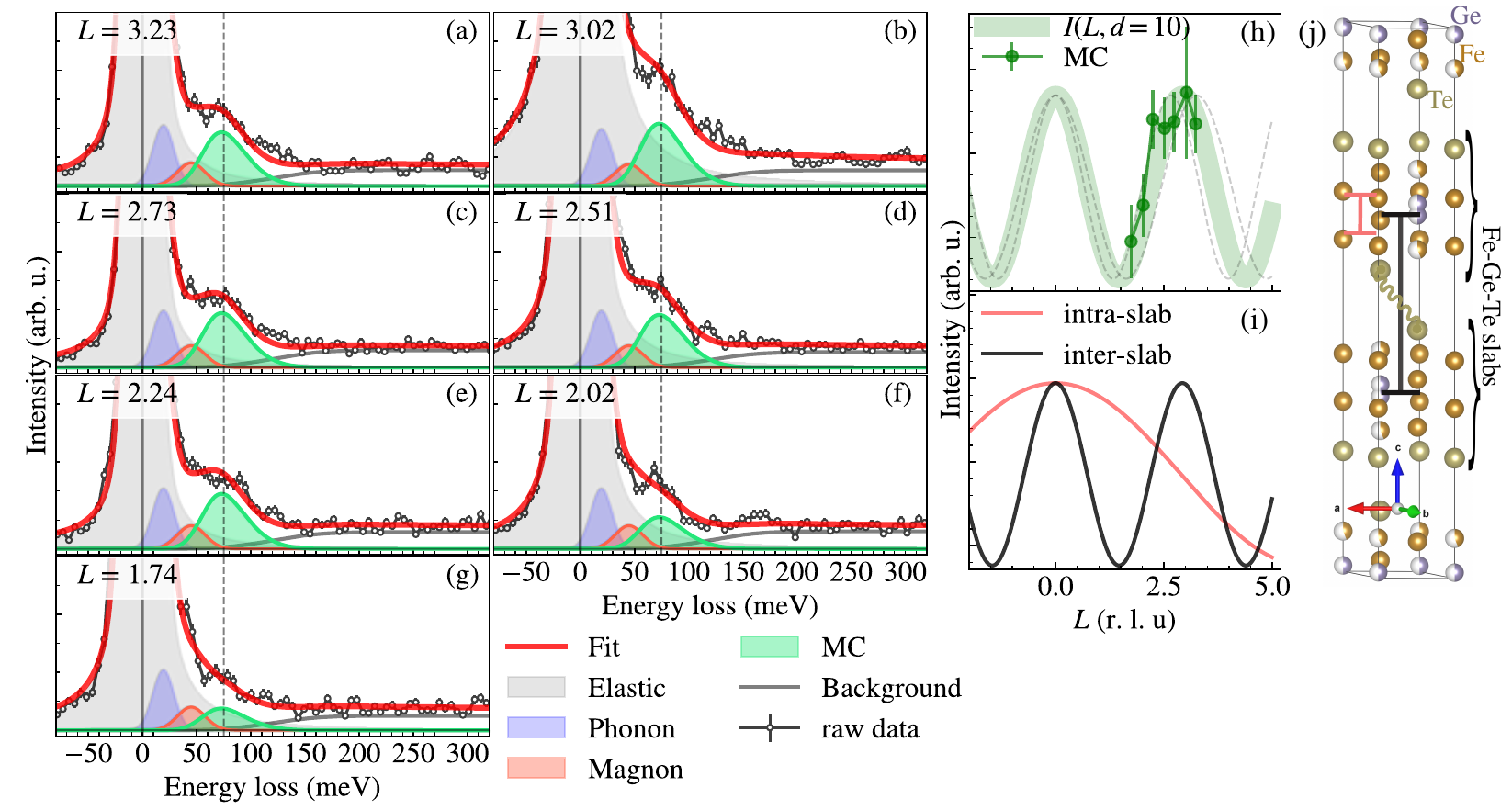}
  \caption{$L$ -- dependent study of the magnetic continuum in \FGT. (a-g) \gls*{RIXS} spectra measured at different $L$ values from 1.74 r.l.u. to 3.23 r.l.u., at 70 K. For all spectra, $H <  |0.04|$ and $K$= 0. Each spectrum is fitted with an elastic peak (gray), a phonon (blue), a magnon (red), and a continuum (green), as introduced in Figs. \ref{FIG:fig1}(e, f). The vertical dashed line marks the continuum centers around $\sim$ 75 meV. (h) $L$-dependence of the continuum integrated spectral weight (green dotted line). The error-bars are half of one standard deviation. 
  The thick green line is a sinusoidal fit of the data using Eq. (\ref{eq:equation}) for $d$ = 10 \AA, while the dotted lines represent the upper and lower boundaries considering the extracted fit error for $d$ of $\pm$ 1 \AA.  (i) Calculated intensity modulation for intra-slab (red) and inter-slab (black) distances, as defined within the crystal structure displayed in (j) using with the same color convention. A gold curved line in (j) highlights the Te -- Te distance across two neighboring slabs.}
  \label{FIG:ldep}
\end{figure*}


The high-quality quenched single crystal of Fe$_{4.75}$GeTe$_2$ used for this study originated from the same batch as in Ref.~\cite{MayPRB2016_Fe3GT,May2019FerromagnetismArticle}, which suggested similar magnetic properties for nominal composition ranging from Fe$_{4.7}$GeTe$_2$ to Fe$_{6.0}$GeTe$_2$. Figure~\ref{FIG:fig1}(a) shows the rhombohedral unit cell of \FGT ($R\bar{3}m$ No. 166, a = b $\approx $ 4.04 \AA, \, c $\approx$ 29.19 \AA), characterized by three distinct \FeGeTe slabs separated by $\approx$10 \AA. Within each slab, the Fe and Ge atoms occupy interior positions, while the Te atoms occupy the external positions. It develops ferromagnetic order below $T_{\rm{C}} \approx$ 315 K  with magnetic moment pointing along the $c-$axis~\cite{May2019FerromagnetismArticle}. 

For the \gls*{XAS} and \gls*{RIXS} studies, the \FGT single crystal was oriented with (1 0 0) and (0 0 1) in the scattering plane, see Fig. \ref{FIG:fig1}(b). The sample was cleaved in air with scotch tape just before loading it into the vacuum chamber. Figure \ref{FIG:fig1}(c) shows the \gls*{XAS} measured in the \gls*{TEY} mode at 70 K. The observed $L_3$ and $L_2$ absorption peaks 
are in agreement with literature \cite{Yamagami2022EnhancedFe5GT}, supporting the high quality of the sample. The \gls*{XAS} data was used to identify the resonant energy for the \gls*{RIXS} study. To reduce the elastic scattering signal in the RIXS measurements, we used $\pi$ polarized incoming x-rays. The same sample was used for both \gls*{REXS} and \gls*{XRD} experiments, and it was aligned using an out-of-plane and two in-plane nuclear reflections. 

The momentum transfer $\bm{Q} (H K L)$ associated with the RIXS spectra and the diffraction data are expressed in  reciprocal lattice units ($4\pi/\sqrt{3}a$ $4\pi/\sqrt{3}a$ $2\pi/c$). The in-plane and out-of-plane momentum coverage was achieved by rotating both the scattering angle $\Omega$ and the x-rays incident angle $\theta$. Extensive momentum coverage was achieved for the \gls*{XRD} experiment (2 keV of photon energy) and for the Fe $K - $edge \gls*{REXS} ( $\approx$ 7 keV of photon energy). For Fe $L_3 - $edge \gls*{RIXS} ($\approx$ 710 eV of photon energy), we could span a large portion of the $\bm{Q}$-space in the $L$ direction across several \gls*{BZ} and up to $L$ = 3.23 $(2\pi/c)$, thanks to the large $c$-axis ($\approx$ 29.19 \AA) of the \FGT, while in-plane we could cover up to $H_{max} = -0.35$ $(4\pi/\sqrt{3}a)$ within the first BZ.

\section{\label{sec:level1}Results and Discussion}

\subsection{Magnetic excitations in \FGT}

In the following two sections we describe the investigation of the magnetic excitations and their momentum dependence in \FGT using Fe $L_3 - $edge RIXS, and we discuss the results in comparison  \FetsGT \cite{Calder2019,Bai2022AntiferromagnetictextGeTe_2, Bao2022Neutron2}. Figure~\ref{FIG:fig1}(d) presents a \gls*{RIXS} intensity map measured as a function of incident photon energy across the Fe $L_3 - $edge. Although most of the spectral weight goes into high-energy fluorescence as expected for a metal (see Supplementary Figure 2) \cite{Pelliciari2021TuningConfinement}, a weak component stretches from 0 up to $\sim$ 150 meV  resonating at the Fe $L_3$ peak. 
By fixing the incident photon energy to the maximum of the Fe $L_3$ resonance, we measured high-resolution RIXS spectra as a function of momentum transfer $\bm{Q}$. 

Figures \ref{FIG:fig1}(e,f) show \gls*{RIXS} spectra 
at two representative $\bm{Q}$ points with largest in-plane momentum components, $\bm{Q}$ = ($-0.35$ 0 1.4) and ($-0.3$ 0 1.74). We make use of both \gls*{RIXS} spectra to establish a minimum fitting model that 
describes the observed inelastic spectral weight below 200 meV. Details regarding the fitting and its procedure are reported in Supplementary Note 3. Taking \FetsGT as a reference case \cite{Bao2022Neutron2,Bai2022AntiferromagnetictextGeTe_2}, we include in the fitting model a resolution limited phonon (blue peak), a  magnon mode (red peak) due to spin-spin correlations between localized spins, and a broad magnetic continuum (green peak) coming from spin-spin interactions among itinerant electrons. 
We fix the phonon energy to 20 meV based on the room temperature Raman measurements presented in Supplementary Figure 1 and consistently with \cite{Bai2022AntiferromagnetictextGeTe_2}. From the remaining spectral weight, we extract a magnon peak centered at 
36 meV $\pm 1$~meV  and a continuum  centered at 
75 meV $\pm 2$~meV,  extending up to $\approx$ 150 meV. At higher energies, a background from the fluorescence tail dominates the spectral weight (gray solid line). \vb{While the magnetic continuum is peaked at twice the energy of the magnon, we exclude a dominant double magnon character, since this type of excitation is approximately 10 times weaker than the magnon~\cite{Li_PRX_2023,Elnaggar_NatComm_2023}. In our dataset, the continuum amplitude is comparable to the magnon, and its area is significantly larger than the latter, ruling out the double magnon character. While we cannot exclude the presence of double magnon in our data, their faint spectral weight would not impact the interpretation nor the extraction of the magnetic continuum. 
}

For the spectra presented in Figs.~\ref{FIG:fig1}(e,\,f), we were not able to resolve any significant magnon dispersion given the proximity of the in-plane $\bm{Q}$ components. 
However, a comparable magnon energy for similar momentum transfer has been reported for \FetsGT \cite{Calder2019, Bai2022AntiferromagnetictextGeTe_2, Bao2022Neutron2}, further supporting the nature of the $\approx$ 36 meV peak. Similarly, the broad continuum peak (stretching up to 150 meV) resembles the one identified in \FethGT \cite{Bao2022Neutron2}, although in the latter system it extends only up to 100 meV. Based on these findings, we identify the low energy magnetic excitations in \FGT, and assess the coexistence of the dual electron spin character in this system.

Our conclusion is underpinned by the \gls*{INS} studies of \FetsGT \cite{Bao2022Neutron2,Bai2022AntiferromagnetictextGeTe_2}. To explain this uncommon behavior in a 3$d$ transition metal \cite{Xie_PRL_2018,Chen_NatComm_2020,Song_npjQM_2021}, an orbital selective Mott transition mechanism has been proposed for \FetsGT  \cite{Bai2022AntiferromagnetictextGeTe_2}, typically invoked in the context of strong multi-orbital  correlations as realized by 4$d$/5$d$ and $f$ electron  compounds \cite{Anisimov2002,Kim_QM_2022, Medici_PRL_2005,Medici_PRL_2009}. Within this scenario, it is possible to explain both the itinerant and the local magnetism manifesting in a multi-orbital system \cite{Bai2022AntiferromagnetictextGeTe_2}. Considering the emergence of giant Fe(3$d$)$-$Te(5$p$) and Fe(3$d$)$-$Ge(4$s/4d$) hybridization \cite{Yamagami2021giant} and the flat bands \cite{Wu2023ReversibleFerromagnet,Wang_PRB_2023}, and the dual character of magnetic excitations evidenced by our study, the magnetic ground state of \FGT is likely governed by \gls*{OSMT} physics as well as in \FetsGT. Theoretical validation would be valuable to confirm this aspect, requiring a comprehensive treatment of the strong multi-orbital correlations manifesting within such a complex crystal structure. 

\subsection{Origin of the magnetic continuum}

Figure \ref{FIG:ldep} shows the $L$ dependent  \gls*{RIXS} spectra along (0 0 $L$) from $L$ = 1.74 r.l.u. to $L$ = 3.23 r.l.u. In these measurements, we maximized the out-of-plane coverage by scanning mostly along (0 0 $L$), while minimizing the in-plane momentum transfer, e.g. $H <  |0.04|$ r.l.u. and $K$ = 0 r.l.u. A clear continuum peak emerges from all spectra, see Figs.\,\ref{FIG:ldep} (a-g), and its position remains mostly invariant versus $L$ as marked by a vertical dashed line. The dispersionless nature of the continuum agrees with the \FetsGT findings \cite{Bao2022Neutron2}. To directly compare the absolute \gls*{RIXS} intensities as a function of $L$ (or scattering geometries), all spectra reported here are normalized to integrated fluorescence peak and self-absorption corrected as described in Supplementary Note 2. 
Already from the raw data, the continuum intensity appears to decrease for decreasing $L$. Note that while in \FetsGT the continuum is maximized around the in-plane \gls*{BZ} boundary~\cite{Bao2022Neutron2}, here we find a significant spectral weight close to the in-plane \gls*{BZ} center.

The origin of the $L$-dependent intensity modulation of the continuum in \FetsGT was associated with the two Fe sites, yielding a characteristic length scale  along the $c-$axis within \FeGeTe slab \cite{Bao2022Neutron2}. The goal of this section is to extract this information for \FGT. We fit the \gls*{RIXS} spectra in Fig. \ref{FIG:ldep} using the model introduced in Figs. \ref{FIG:fig1}(e,\,f). 
To effectively implement the fitting model, we made some key assumptions. 
Given the strong elastic line intensity and the unknown magnon dispersion for this material, we fixed the fitting parameters (peak position and width) of the lowest energy modes - phonon and magnon - to match the values extracted for $\bm{Q}$ = ($-0.35$ 0 1.4) r.l.u. 
\vb{These assumptions are valid because the low-energy spectral weight (below 40 meV) only influences the elastic line shape and leaves the high-energy dispersionless continuum unaffected. This approach enabled us to utilize a consistent fitting model across all RIXS spectra.
}
Regarding the fitting of the continuum itself, we fixed its position and width to reproduce the broad and dispersionless character of this excitation \cite{Bao2022Neutron2,Bai2022AntiferromagnetictextGeTe_2}, while its amplitude was allowed to change. More details on the fitting procedure can be found in Supplementary Note 3. 
As a result of this analysis, we obtain the $L$-dependent evolution of the continuum intensity  which follows a sinusoidal behavior, as highlighted in Fig. \ref{FIG:ldep} (h).  A comparison of the $L$ - dependent magnetic continuum intensity with and without self-absorption correction is shown in Supplementary Figure 3. The magnitude of intensity modulation is robust and independent of self-absorption correction.

When dealing with a continuum or a  dispersionless excitation, the $\bm{Q}$-dependent modulation of its intensity or structure factor encodes the distance between the interacting atoms along the selected $\bm{Q}$-direction. Hard x-ray RIXS and INS -- enabling a wide coverage of  reciprocal space -- have been extensively used to show the formation of new magnetic superstructures by monitoring the $\bm{Q}$-modulation of the spectral weight  
~\cite{Revelli2020FingerprintsIridates,Revelli2022Quasimolecular, Bao2022Neutron2}. Following these works, the frequency of the continuum structure factor modulation can be captured by a simple qualitative function:

\begin{equation} \label{eq:equation}
    I(L) \propto \cos{(L \cdot {d}/2)}^2,
\end{equation}
where $L$ is the momentum transfer vector and ${d}$ is the real space inter-atomic distance along the $c-$axis within the unit cell. By fitting the data points displayed in Fig. \ref{FIG:ldep} (h), we extract a distance of  ${d}$ = $ 10 \pm 1$ \AA. 

In the case of \FetsGT, the $L$-dependent INS structure factor of the continuum yielded an intra-slab distance ${d} \sim 2.7$ \AA \,
\cite{Bao2022Neutron2}. Such a length-scale corresponds to the Fe-Fe dumbbell (a dimer like structure corresponding to the largest exchange interaction) superstructure oriented along the $c-$axis within an individual \FeGeTe slab. This finding suggests that the magnetic continuum in \FetsGT originates from the quasi-molecular character of the magnetic states formed by the Fe-Fe dumbbell, and furthermore it identifies the atoms and exchange path that contribute to the strongest magnetic exchange interaction in the system~\cite{Bai2022AntiferromagnetictextGeTe_2}.

To identify the interacting atoms and exchange path behind the continuum intensity modulation in \FGT, we display in Fig. \ref{FIG:ldep} (i) a simulation of $I(L)$ based on a representative intra-slab distance (red), 
and inter-slab distance (black). Such distances are highlighted in the unit cell of Fig. \ref{FIG:ldep} (l), using the same color code. The intra-slab distance $d \sim 2.53$ \AA\, connecting two Fe atoms within the same \FeGeTe slab fails in reproducing the observed modulation, despite it well described the $L$-dependent continuum modulation for \FetsGT \cite{Bao2022Neutron2}. Rather, we find that an inter-slab distance $d \sim 11$ \AA \, connecting the center of two consecutive slabs well reproduces the continuum modulation for \FGT. Our finding supports the presence of an inter-slab exchange path along the $c$-axis, suggesting a three-dimensional character for the magnetic interaction in \FGT. 
Additionally, we identify the \FeGeTe slabs as the interacting magnetic units in \FGT. This is consistent with recent \gls*{XMCD} and \gls*{ARPES} studies underlining the key role of Te and Ge in determining the itinerant long-range ferromagnetism in \FGT \cite{Yamagami2022EnhancedFe5GT, Yamagami2021giant}, owing to the strong hybridization and covalent nature of the Fe-Te and the Fe-Ge bonds. Based on atomic position considerations, we suggest that the shorter Fe-Te nearest neighbor distance (2.63 \AA ) in \FGT with respect to \FetsGT (2.643 \AA) possibly favors the inter-slab exchange mediated by the Te sites (highlighted by the curved line in Fig. \ref{FIG:ldep}(j)). Overall, we speculate that the inter-slab exchange interaction revealed by our study could be crucial for determining the high $T_{\rm C}$ in \FGT, endorsed by recent studies on epitaxially grown thin films where $T_{\rm C}$ is found to be inversely proportional to the inter-slab distance~\cite{SilinskasOnFilms}.

\begin{figure*}
  \includegraphics[width=1\textwidth]{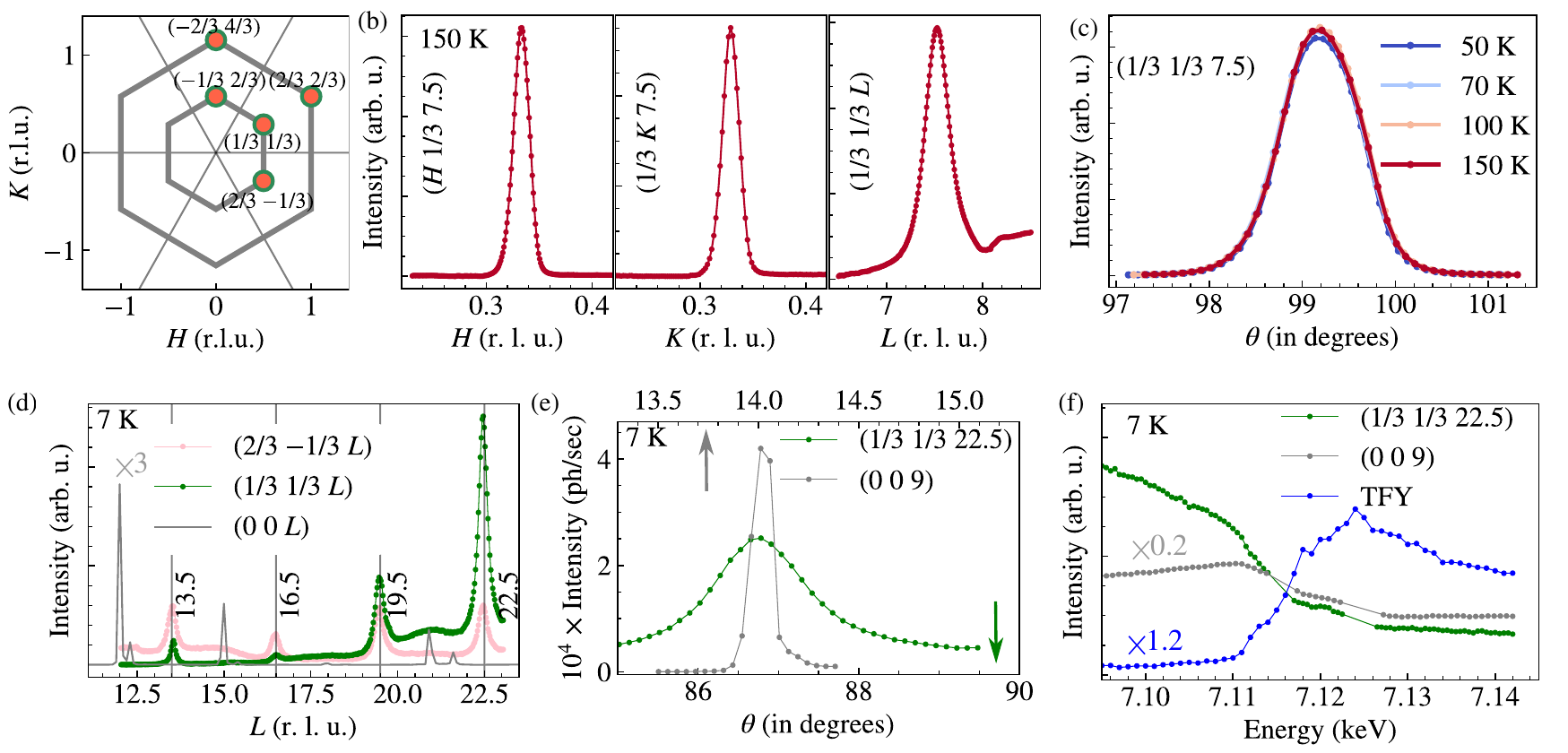}
  \caption{Investigation of charge order in \FGT by \gls*{REXS} and \gls*{XRD}. (a) A sketch highlighting the measured symmetry equivalent $\bm{Q}$ positions. (b) $H$, $K$, and $L$ scans of a representative (1/3 1/3 7.5) peak using 2.1 keV x-ray energy at 150 K. (c) The temperature dependence  of the (1/3 1/3 7.5) peak.
  (d) Fixed-energy $\bm{Q}$-scan of (1/3 1/3 $L$) (green dotted line), (2/3 -1/3 $L$) (pink dotted line), and of (0 0 $L$) (gray line) along the $L$ direction. The temperature was set to 7 K for this measurement and the x-ray energy was tuned prior to the Fe $K - $edge, $\sim$ 7.1 keV. Double peaks in (0 0 $L$) scan are due to a second crystallite. (0 0 $L$) scan is rescaled to match y-scale. (e) Absolute intensity comparison between (1/3 1/3 22.5) and (0 0 9) peaks. (f) Fixed-$\bm{Q}$ energy scan of (1/3 1/3 22.5) (green dotted line) and (0 0 9) (gray dotted line) peaks, rescaled to match y-scale. The Fe $K - $edge \gls*{XAS} of \FGT measured in total fluoresce yield (TFY) mode is displayed (rescaled) for reference (blue dotted line).}\label{FIG:charge_order}
\end{figure*}


\subsection{Absence of charge order in bulk \FGT} \label{sec: charge order}
Recently, selected area electron diffraction studies \cite{Gao2020SpontaneousFe5xGeTe2} reported a superstructure modulation vector $\bm{Q}_1$= $\pm $1/3 (1 1 3) above 100 K, fading away at lower temperatures with a diffused line along 1/3 (1 1 $L$) together with a new $\bm{Q}_2$ = $\pm $3/10 (0  0  3$L$) order. A short range $\sqrt{3}a \times \sqrt{3}a$ $R 30^\circ$ order was observed by single crystal \gls*{XRD} and the high-angle annular dark field image and associated with Fe occupancy of a crystallographic split-site \cite{May2019FerromagnetismArticle}. Further STM and ARPES investigations \cite{Wu2021DirectGeTe_2} confirmed the existence of a $\sqrt{3}a \times \sqrt{3}a$ $R 30^\circ$ in-plane periodic modulation and assigned it to the charge ordering. To further the understanding of the bulk signature, electronic character and propagation vectors of $\bm{Q}_1$ and $\bm{Q}_2$ in \FGT, in the following, we present synchrotron \gls*{XRD} ($\sim$ 2.1 keV) and \gls*{REXS} (Fe $K - $edge, $\sim$ 7.1 keV)  investigations. 

We confirmed the in-plane order $\sqrt{3}a \times \sqrt{3}a$ $R 30^\circ$ in  our \FGT single crystal by finding several symmetry-equivalent peaks at 150 K, as represented in Fig. \ref{FIG:charge_order}(a). Figure  \ref{FIG:charge_order}(b) reports the $H$, $K$, and $L$ scans of a representative peak with (1/3 1/3) in-plane components, measured using $\sim$ 2.1 keV x-rays. Interestingly, we find that the out-of-plane component peaks at  $L$ = 7.5 r.l.u. No significant temperature dependence is detected for this peak while going across 100 K, as displayed in Fig. \ref{FIG:charge_order}(c). The extended $L$ scan of (2/3 -1/3 $L$) and (1/3 1/3 $L$) peaks are presented in Fig. \ref{FIG:charge_order}(d), using $\sim$ 7.1 keV x-rays (dotted lines): a clear (1/3 1/3 3$n$+1.5) with $n$ = 0, 1, 2, 3.... periodicity emerges even at 7 K.  
The observed (1/3 1/3 3$n$+1.5) modulation differs from  the previously reported $\pm $1/3 (1 1 3) order \cite{Gao2020SpontaneousFe5xGeTe2} and it is also distinct from the (0 0 3$n$) selection rule expected for a rhombohedral  unit cell (space group $R\bar{3}m$). For reference, we reproduce in Fig.~\ref{FIG:charge_order}(d) the (0 0 $L$) scan (gray line) showing the (0 0 3$n$) periodicity of the structural Bragg peak. The satellite structure visible in the (0 0 $L$) scan is attributed to a second crystallite present in the sample. Also no fingerprint of $\bm{Q}_2$ =  $\pm $3/10(0  0  3$L$)   is observed in our data down to 7 K. 

The intensities of (0 0 $L$) and (2/3 -1/3 $L$) or (1/3 1/3 $L$) scans cannot be directly compared in Fig.~\ref{FIG:charge_order}(d), as we used a higher intensity detuning filter for the main structural Bragg peaks along (0 0 $L$). An absolute intensity comparison between (1/3 1/3 22.5) and (0 0 9) is presented in \ref{FIG:charge_order}(\vb{e}), showing that the intensities of both peaks are of the same order of magnitude, with the (0 0 9) peak approximately 50\% stronger than (1/3 1/3 22.5).

Additionally, using Fe $K - $edge \gls*{REXS}, we investigated the resonance effect of the (1/3 1/3 22.5) peak. The resulting fixed-$\bm{Q}$ energy scan is presented in Fig. \ref{FIG:charge_order}(e) (green dotted line), and compared to the energy scan of the (0 0 9) structural Bragg peak (gray line). Both scans present a similar behavior, with no resonance enhancement. The Fe $K - $edge XAS of \FGT is reproduced for reference in the same figure.

Overall, our \gls*{REXS} and \gls*{XRD} studies  report the presence of a
commensurate (1/3 1/3 $L$) Bragg peak with a 3$n$+1.5 out-of-plane modulation, no temperature dependence down to 7 K, and no resonance effect. Based on these observations, we propose these peaks  originate from the crystal structure rather than  charge order \,
\cite{Abbamonte2004CrystallizationSr14Cu24O41,Chen2016Remarkable4,Shen2021ChargeNickelates, Shan2023chargeorder}. This conclusion is further corroborated by the comparable peak intensity with respect to main structural Bragg peaks. Furthermore, we underline that the observed 3$n$+1.5 out-of-plane periodicity is not compatible with the space group $R\bar{3}m$ proposed for \FGT, instead, it infers doubling of structural unit cell along the $\bm{c}$-axis and reduced crystal symmetry. One direct consequence of reduced lattice symmetry would be a more complex network of magnetic interactions, which might be crucial to interpret the various kinks in the magnetization versus temperature diagram \cite{May2019FerromagnetismArticle}.

To reconcile the discrepancy between our observations and the ones reported in the literature \cite{Gao2020SpontaneousFe5xGeTe2, Wu2021DirectGeTe_2}, we should acknowledge intrinsic differences connected with the techniques involved. STM and ARPES mostly probe the surface, which in certain materials can significantly differ from the bulk \cite{Brown2005SurfaceTransition, Gu2023DetectionUTe2, Kengle2024Absence2,Theuss2024Absence2}. Electron diffraction instead probes the bulk, but the required sample preparation methods - such as focused ion beam and grinding - might introduce permanent micro-structural changes affecting the material response \cite{Mayer2007TEMDamage, Volkert2007FocusedMicromachining}. Finally, we cannot exclude a sample to sample variation in determining the ultimate material properties. \FGT single crystals are known to have stacking faults and different samples (especially grown by different groups) might even have slightly different Fe concentration. To advance the understanding of this complex system, it would be helpful to perform on the same sample both surface sensitive (STM, ARPES) and bulk sensitive (XRD, REXS) investigations.


In summary, our work sheds light on the magnetic excitations and electronic order in the room temperature vdW ferromagnet \FGT. Using high-resolution RIXS at the Fe $L_3 - $edge, we reveal that \FGT hosts dual character of magnetic excitations composed of a magnon and a continuum, similarly to the more extensively characterized \FetsGT, and prove that the dual character of magnetic excitation, known to be driven by orbital selective Mott transition, is common to ternary \FeGeTe vdW materials.  The out-of-plane modulation of the continuum intensity reveals the presence of a strong out-of-plane inter-slab exchange path, where the interacting magnetic units are the neighboring \FeGeTe slabs. We conclude that \FGT behaves like a 3D magnet, and we discuss this finding in connection to its $T_{\rm C}$, the highest among \FeGeTe vdW materials. 
Additionally, employing \gls*{XRD} and \gls*{REXS}, we explored the bulk character, the electronic origin and the propagation vector of the previously reported charge order with (1/3 1/3) in-plane components.  The absence of temperature and resonance effects, and the commensurability with the crystal structure led us to conclude that, the observed Bragg peaks do not originate from charge order but rather  crystal structure with reduced symmetry. The presence of the (1/3 1/3 3$n$+1.5) peaks suggests a  doubling of the unit cell along the c-axis, indicating a subtle change in the stacking sequence or atomic positions that is not captured by the $R\bar{3}m$ space group. 

\section*{Methods}
\subsection*{X-ray Absorption (XAS) and resonant inelastic x-ray Scattering (RIXS)}
High-resolution \gls*{XAS} and \gls*{RIXS} experiments were carried out at the SIX 2-ID beamline of NSLS-II ~\cite{DovrakRSI2016}. The experimental energy resolution at the Fe $L_3 - $edge ($\sim$ 708 eV) was 21 meV for the RIXS measurements, determined by the full width at half-maximum of the elastic peak measured from a reference multilayer sample. To suppress the elastic line, $\pi-$ polarization was used throughout the experiment.
 
\subsection*{Resonant elastic x-ray scattering (REXS) and x-ray diffraction (XRD)}
Synchrotron based \gls*{XRD} was used to measure the rocking curves and temperature dependence of (1/3 1/3 7.5) peak at 29-ID beamline of APS, with a photon energy of 2.1 keV. The Fe $K-$ edge \gls*{REXS} was performed at the 4-ID beamline of NSLS-II.

\subsection*{Raman Spectroscopy}
We performed room-temperature Raman scattering ($\lambda$ = 6328 \AA) on a \FGT flake exfoliated on SiO$_2$/Si and oriented as $c$-axis normal using a LabRAM HR Evolution Raman microscope from Horiba Scientific using 100 $\times$ 0.9NA Olympus air objective. Spectra were recorded using an 1800 gr/mm holographic grating and Synapse II front-illuminated EMCCD (Supplementary Figure 1). 
\section*{Data availability}
Data supporting the findings of this study are available from the corresponding author on reasonable request.

\section*{Acknowledgments}
This work was supported as part of Programmable Quantum Materials, an Energy Frontier Research Center funded by the US Department of Energy (DOE), Office of Science, Basic Energy Sciences (BES), under award DE-SC0019443. 
Development of \FGT crystals (AFM) was supported by the U. S. Department of Energy, Office of Science, Basic Energy Sciences, Materials Sciences and Engineering Division.
This research used beamlines 2-ID and 4-ID of NSLS-II, a US DOE Office of Science User Facility operated for the DOE Office of Science by Brookhaven National Laboratory under contract no. DE-SC0012704. This research used resources of the Advanced Photon Source, a U.S. Department of Energy (DOE) Office of Science user facility operated for the DOE Office of Science by Argonne National Laboratory under Contract No. DE-AC02-06CH11357.

\section*{Author Contributions}
VBi, VBh, CM, AM, DB, and XX conceived the research plan. AFM grew and characterized the bulk single crystalline \FGT. VBi, VBh, TK, JL, YG, SF, JP  carried out the RIXS measurements. VBh, TK, CM, CN, JF performed the \gls*{XRD} and \gls*{REXS} experiments. TD and DJR realized the Raman measurements. VBh and VBi analyzed and interpreted the data, with contributions from all coauthors. VBh and VBi wrote the manuscript. All authors commented on the text.

\section{Competing Interests}
The authors declare no competing interests.

\bibliography{references_FGT}

\begin{thebibliography}{3}%
\makeatletter
\providecommand \@ifxundefined [1]{%
 \@ifx{#1\undefined}
}%
\providecommand \@ifnum [1]{%
 \ifnum #1\expandafter \@firstoftwo
 \else \expandafter \@secondoftwo
 \fi
}%
\providecommand \@ifx [1]{%
 \ifx #1\expandafter \@firstoftwo
 \else \expandafter \@secondoftwo
 \fi
}%
\providecommand \natexlab [1]{#1}%
\providecommand \enquote  [1]{``#1''}%
\providecommand \bibnamefont  [1]{#1}%
\providecommand \bibfnamefont [1]{#1}%
\providecommand \citenamefont [1]{#1}%
\providecommand \href@noop [0]{\@secondoftwo}%
\providecommand \href [0]{\begingroup \@sanitize@url \@href}%
\providecommand \@href[1]{\@@startlink{#1}\@@href}%
\providecommand \@@href[1]{\endgroup#1\@@endlink}%
\providecommand \@sanitize@url [0]{\catcode `\\12\catcode `\$12\catcode
  `\&12\catcode `\#12\catcode `\^12\catcode `\_12\catcode `\%12\relax}%
\providecommand \@@startlink[1]{}%
\providecommand \@@endlink[0]{}%
\providecommand \url  [0]{\begingroup\@sanitize@url \@url }%
\providecommand \@url [1]{\endgroup\@href {#1}{\urlprefix }}%
\providecommand \urlprefix  [0]{URL }%
\providecommand \Eprint [0]{\href }%
\providecommand \doibase [0]{https://doi.org/}%
\providecommand \selectlanguage [0]{\@gobble}%
\providecommand \bibinfo  [0]{\@secondoftwo}%
\providecommand \bibfield  [0]{\@secondoftwo}%
\providecommand \translation [1]{[#1]}%
\providecommand \BibitemOpen [0]{}%
\providecommand \bibitemStop [0]{}%
\providecommand \bibitemNoStop [0]{.\EOS\space}%
\providecommand \EOS [0]{\spacefactor3000\relax}%
\providecommand \BibitemShut  [1]{\csname bibitem#1\endcsname}%
\let\auto@bib@innerbib\@empty
\bibitem [{\citenamefont {Wang}\ \emph {et~al.}(2020)\citenamefont {Wang},
  \citenamefont {Elnaggar}, \citenamefont {Titus}, \citenamefont {Tomiyasu},
  \citenamefont {Geessinck}, \citenamefont {Koster}, \citenamefont {Frati},
  \citenamefont {Okamoto}, \citenamefont {Huang},\ and\ \citenamefont
  {De~Groot}}]{Wang_Saturation_2020}%
  \BibitemOpen
  \bibfield  {author} {\bibinfo {author} {\bibfnamefont {R.-P.}\ \bibnamefont
  {Wang}}, \bibinfo {author} {\bibfnamefont {H.}~\bibnamefont {Elnaggar}},
  \bibinfo {author} {\bibfnamefont {C.~J.}\ \bibnamefont {Titus}}, \bibinfo
  {author} {\bibfnamefont {K.}~\bibnamefont {Tomiyasu}}, \bibinfo {author}
  {\bibfnamefont {J.}~\bibnamefont {Geessinck}}, \bibinfo {author}
  {\bibfnamefont {G.}~\bibnamefont {Koster}}, \bibinfo {author} {\bibfnamefont
  {F.}~\bibnamefont {Frati}}, \bibinfo {author} {\bibfnamefont
  {J.}~\bibnamefont {Okamoto}}, \bibinfo {author} {\bibfnamefont {D.-J.}\
  \bibnamefont {Huang}},\ and\ \bibinfo {author} {\bibfnamefont {F.~M.~F.}\
  \bibnamefont {De~Groot}},\ }\bibfield  {title} {\bibinfo {title} {{Saturation
  and self-absorption effects in the angle-dependent 2p3d resonant inelastic
  X-ray scattering spectra of Co$^{3+}$}},\ }\href
  {https://doi.org/10.1107/S1600577520005123} {\bibfield  {journal} {\bibinfo
  {journal} {J. Synchrotron Rad}\ }\textbf {\bibinfo {volume} {27}},\ \bibinfo
  {pages} {979} (\bibinfo {year} {2020})}\BibitemShut {NoStop}%
\bibitem [{\citenamefont {Robarts}\ \emph {et~al.}(2021)\citenamefont
  {Robarts}, \citenamefont {Garc\'{\i}a-Fern\'andez}, \citenamefont {Li},
  \citenamefont {Nag}, \citenamefont {Walters}, \citenamefont {Headings},
  \citenamefont {Hayden},\ and\ \citenamefont
  {Zhou}}]{Robarts2021DynamicalScattering}%
  \BibitemOpen
  \bibfield  {author} {\bibinfo {author} {\bibfnamefont {H.~C.}\ \bibnamefont
  {Robarts}}, \bibinfo {author} {\bibfnamefont {M.}~\bibnamefont
  {Garc\'{\i}a-Fern\'andez}}, \bibinfo {author} {\bibfnamefont
  {J.}~\bibnamefont {Li}}, \bibinfo {author} {\bibfnamefont {A.}~\bibnamefont
  {Nag}}, \bibinfo {author} {\bibfnamefont {A.~C.}\ \bibnamefont {Walters}},
  \bibinfo {author} {\bibfnamefont {N.~E.}\ \bibnamefont {Headings}}, \bibinfo
  {author} {\bibfnamefont {S.~M.}\ \bibnamefont {Hayden}},\ and\ \bibinfo
  {author} {\bibfnamefont {K.-J.}\ \bibnamefont {Zhou}},\ }\bibfield  {title}
  {\bibinfo {title} {Dynamical spin susceptibility in {La}$_{2}${CuO}$_{4}$
  studied by resonant inelastic x-ray scattering},\ }\href
  {https://doi.org/10.1103/PhysRevB.103.224427} {\bibfield  {journal} {\bibinfo
   {journal} {Phys. Rev. B}\ }\textbf {\bibinfo {volume} {103}},\ \bibinfo
  {pages} {224427} (\bibinfo {year} {2021})}\BibitemShut {NoStop}%
\bibitem [{\citenamefont {Lin}\ \emph {et~al.}(2020)\citenamefont {Lin},
  \citenamefont {Miao}, \citenamefont {Mazzone}, \citenamefont {Gu},
  \citenamefont {Nag}, \citenamefont {Walters}, \citenamefont
  {Garc\'{\i}a-Fern\'andez}, \citenamefont {Barbour}, \citenamefont
  {Pelliciari}, \citenamefont {Jarrige}, \citenamefont {Oda}, \citenamefont
  {Kurosawa}, \citenamefont {Momono}, \citenamefont {Zhou}, \citenamefont
  {Bisogni}, \citenamefont {Liu},\ and\ \citenamefont
  {Dean}}]{Lin2070StronglyAnomaly}%
  \BibitemOpen
  \bibfield  {author} {\bibinfo {author} {\bibfnamefont {J.~Q.}\ \bibnamefont
  {Lin}}, \bibinfo {author} {\bibfnamefont {H.}~\bibnamefont {Miao}}, \bibinfo
  {author} {\bibfnamefont {D.~G.}\ \bibnamefont {Mazzone}}, \bibinfo {author}
  {\bibfnamefont {G.~D.}\ \bibnamefont {Gu}}, \bibinfo {author} {\bibfnamefont
  {A.}~\bibnamefont {Nag}}, \bibinfo {author} {\bibfnamefont {A.~C.}\
  \bibnamefont {Walters}}, \bibinfo {author} {\bibfnamefont {M.}~\bibnamefont
  {Garc\'{\i}a-Fern\'andez}}, \bibinfo {author} {\bibfnamefont
  {A.}~\bibnamefont {Barbour}}, \bibinfo {author} {\bibfnamefont
  {J.}~\bibnamefont {Pelliciari}}, \bibinfo {author} {\bibfnamefont
  {I.}~\bibnamefont {Jarrige}}, \bibinfo {author} {\bibfnamefont
  {M.}~\bibnamefont {Oda}}, \bibinfo {author} {\bibfnamefont {K.}~\bibnamefont
  {Kurosawa}}, \bibinfo {author} {\bibfnamefont {N.}~\bibnamefont {Momono}},
  \bibinfo {author} {\bibfnamefont {K.-J.}\ \bibnamefont {Zhou}}, \bibinfo
  {author} {\bibfnamefont {V.}~\bibnamefont {Bisogni}}, \bibinfo {author}
  {\bibfnamefont {X.}~\bibnamefont {Liu}},\ and\ \bibinfo {author}
  {\bibfnamefont {M.~P.~M.}\ \bibnamefont {Dean}},\ }\bibfield  {title}
  {\bibinfo {title} {Strongly correlated charge density wave in
  {La}$_{2-x}${Sr}$_{x}${CuO}$_{4}$ evidenced by doping-dependent phonon
  anomaly},\ }\href {https://doi.org/10.1103/PhysRevLett.124.207005} {\bibfield
   {journal} {\bibinfo  {journal} {Phys. Rev. Lett.}\ }\textbf {\bibinfo
  {volume} {124}},\ \bibinfo {pages} {207005} (\bibinfo {year}
  {2020})}\BibitemShut {NoStop}%
\end{thebibliography}%


\begin{thebibliography}{54}%
\makeatletter
\providecommand \@ifxundefined [1]{%
 \@ifx{#1\undefined}
}%
\providecommand \@ifnum [1]{%
 \ifnum #1\expandafter \@firstoftwo
 \else \expandafter \@secondoftwo
 \fi
}%
\providecommand \@ifx [1]{%
 \ifx #1\expandafter \@firstoftwo
 \else \expandafter \@secondoftwo
 \fi
}%
\providecommand \natexlab [1]{#1}%
\providecommand \enquote  [1]{``#1''}%
\providecommand \bibnamefont  [1]{#1}%
\providecommand \bibfnamefont [1]{#1}%
\providecommand \citenamefont [1]{#1}%
\providecommand \href@noop [0]{\@secondoftwo}%
\providecommand \href [0]{\begingroup \@sanitize@url \@href}%
\providecommand \@href[1]{\@@startlink{#1}\@@href}%
\providecommand \@@href[1]{\endgroup#1\@@endlink}%
\providecommand \@sanitize@url [0]{\catcode `\\12\catcode `\$12\catcode
  `\&12\catcode `\#12\catcode `\^12\catcode `\_12\catcode `\%12\relax}%
\providecommand \@@startlink[1]{}%
\providecommand \@@endlink[0]{}%
\providecommand \url  [0]{\begingroup\@sanitize@url \@url }%
\providecommand \@url [1]{\endgroup\@href {#1}{\urlprefix }}%
\providecommand \urlprefix  [0]{URL }%
\providecommand \Eprint [0]{\href }%
\providecommand \doibase [0]{https://doi.org/}%
\providecommand \selectlanguage [0]{\@gobble}%
\providecommand \bibinfo  [0]{\@secondoftwo}%
\providecommand \bibfield  [0]{\@secondoftwo}%
\providecommand \translation [1]{[#1]}%
\providecommand \BibitemOpen [0]{}%
\providecommand \bibitemStop [0]{}%
\providecommand \bibitemNoStop [0]{.\EOS\space}%
\providecommand \EOS [0]{\spacefactor3000\relax}%
\providecommand \BibitemShut  [1]{\csname bibitem#1\endcsname}%
\let\auto@bib@innerbib\@empty
\bibitem [{\citenamefont {Burch}\ \emph {et~al.}(2018)\citenamefont {Burch},
  \citenamefont {Mandrus},\ and\ \citenamefont
  {Park}}]{Burch2018MagnetismMaterials}%
  \BibitemOpen
  \bibfield  {author} {\bibinfo {author} {\bibfnamefont {K.~S.}\ \bibnamefont
  {Burch}}, \bibinfo {author} {\bibfnamefont {D.}~\bibnamefont {Mandrus}},\
  and\ \bibinfo {author} {\bibfnamefont {J.~G.}\ \bibnamefont {Park}},\
  }\bibfield  {title} {\bibinfo {title} {{Magnetism in two-dimensional van der
  Waals materials}},\ }\href {https://doi.org/10.1038/s41586-018-0631-z}
  {\bibfield  {journal} {\bibinfo  {journal} {Nature}\ }\textbf {\bibinfo
  {volume} {563}},\ \bibinfo {pages} {47} (\bibinfo {year} {2018})}\BibitemShut
  {NoStop}%
\bibitem [{\citenamefont {Gibertini}\ \emph {et~al.}(2019)\citenamefont
  {Gibertini}, \citenamefont {Koperski}, \citenamefont {Morpurgo},\ and\
  \citenamefont {Novoselov}}]{Gibertini2019MagneticHeterostructures}%
  \BibitemOpen
  \bibfield  {author} {\bibinfo {author} {\bibfnamefont {M.}~\bibnamefont
  {Gibertini}}, \bibinfo {author} {\bibfnamefont {M.}~\bibnamefont {Koperski}},
  \bibinfo {author} {\bibfnamefont {A.~F.}\ \bibnamefont {Morpurgo}},\ and\
  \bibinfo {author} {\bibfnamefont {K.~S.}\ \bibnamefont {Novoselov}},\
  }\bibfield  {title} {\bibinfo {title} {Magnetic {2D} materials and
  heterostructures},\ }\href {https://doi.org/10.1038/s41565-019-0438-6}
  {\bibfield  {journal} {\bibinfo  {journal} {Nature Nanotechnology}\ }\textbf
  {\bibinfo {volume} {14}},\ \bibinfo {pages} {408} (\bibinfo {year}
  {2019})}\BibitemShut {NoStop}%
\bibitem [{\citenamefont {Wang}\ \emph {et~al.}(2021)\citenamefont {Wang},
  \citenamefont {Bedoya-Pinto}, \citenamefont {Blei}, \citenamefont {Dismukes},
  \citenamefont {Hamo}, \citenamefont {Jenkins}, \citenamefont {Koperski},
  \citenamefont {Liu}, \citenamefont {Sun}, \citenamefont {Telford},
  \citenamefont {Kim}, \citenamefont {Augustin}, \citenamefont {Vool},
  \citenamefont {Yin}, \citenamefont {Li}, \citenamefont {Falin}, \citenamefont
  {Dean}, \citenamefont {Casanova}, \citenamefont {Evans}, \citenamefont
  {Chshiev}, \citenamefont {Mishchenko}, \citenamefont {Petrovic},
  \citenamefont {He}, \citenamefont {Zhao}, \citenamefont {Tsen}, \citenamefont
  {Gerardot}, \citenamefont {Brotons-Gisbert}, \citenamefont {Guguchia},
  \citenamefont {Roy}, \citenamefont {Tongay}, \citenamefont {Wang},
  \citenamefont {Hasan}, \citenamefont {Wrachtrup}, \citenamefont {Yacoby},
  \citenamefont {Fert}, \citenamefont {Parkin}, \citenamefont {Novoselov},
  \citenamefont {Dai}, \citenamefont {Balicas},\ and\ \citenamefont
  {Santos}}]{Wang2021TheMaterials}%
  \BibitemOpen
  \bibfield  {author} {\bibinfo {author} {\bibfnamefont {Q.~H.}\ \bibnamefont
  {Wang}}, \bibinfo {author} {\bibfnamefont {A.}~\bibnamefont {Bedoya-Pinto}},
  \bibinfo {author} {\bibfnamefont {M.}~\bibnamefont {Blei}}, \bibinfo {author}
  {\bibfnamefont {A.~H.}\ \bibnamefont {Dismukes}}, \bibinfo {author}
  {\bibfnamefont {A.}~\bibnamefont {Hamo}}, \bibinfo {author} {\bibfnamefont
  {S.}~\bibnamefont {Jenkins}}, \bibinfo {author} {\bibfnamefont
  {M.}~\bibnamefont {Koperski}}, \bibinfo {author} {\bibfnamefont
  {Y.}~\bibnamefont {Liu}}, \bibinfo {author} {\bibfnamefont {Q.~C.}\
  \bibnamefont {Sun}}, \bibinfo {author} {\bibfnamefont {E.~J.}\ \bibnamefont
  {Telford}}, \bibinfo {author} {\bibfnamefont {H.~H.}\ \bibnamefont {Kim}},
  \bibinfo {author} {\bibfnamefont {M.}~\bibnamefont {Augustin}}, \bibinfo
  {author} {\bibfnamefont {U.}~\bibnamefont {Vool}}, \bibinfo {author}
  {\bibfnamefont {J.~X.}\ \bibnamefont {Yin}}, \bibinfo {author} {\bibfnamefont
  {L.~H.}\ \bibnamefont {Li}}, \bibinfo {author} {\bibfnamefont
  {A.}~\bibnamefont {Falin}}, \bibinfo {author} {\bibfnamefont {C.~R.}\
  \bibnamefont {Dean}}, \bibinfo {author} {\bibfnamefont {F.}~\bibnamefont
  {Casanova}}, \bibinfo {author} {\bibfnamefont {R.~F.}\ \bibnamefont {Evans}},
  \bibinfo {author} {\bibfnamefont {M.}~\bibnamefont {Chshiev}}, \bibinfo
  {author} {\bibfnamefont {A.}~\bibnamefont {Mishchenko}}, \bibinfo {author}
  {\bibfnamefont {C.}~\bibnamefont {Petrovic}}, \bibinfo {author}
  {\bibfnamefont {R.}~\bibnamefont {He}}, \bibinfo {author} {\bibfnamefont
  {L.}~\bibnamefont {Zhao}}, \bibinfo {author} {\bibfnamefont {A.~W.}\
  \bibnamefont {Tsen}}, \bibinfo {author} {\bibfnamefont {B.~D.}\ \bibnamefont
  {Gerardot}}, \bibinfo {author} {\bibfnamefont {M.}~\bibnamefont
  {Brotons-Gisbert}}, \bibinfo {author} {\bibfnamefont {Z.}~\bibnamefont
  {Guguchia}}, \bibinfo {author} {\bibfnamefont {X.}~\bibnamefont {Roy}},
  \bibinfo {author} {\bibfnamefont {S.}~\bibnamefont {Tongay}}, \bibinfo
  {author} {\bibfnamefont {Z.}~\bibnamefont {Wang}}, \bibinfo {author}
  {\bibfnamefont {M.~Z.}\ \bibnamefont {Hasan}}, \bibinfo {author}
  {\bibfnamefont {J.}~\bibnamefont {Wrachtrup}}, \bibinfo {author}
  {\bibfnamefont {A.}~\bibnamefont {Yacoby}}, \bibinfo {author} {\bibfnamefont
  {A.}~\bibnamefont {Fert}}, \bibinfo {author} {\bibfnamefont {S.}~\bibnamefont
  {Parkin}}, \bibinfo {author} {\bibfnamefont {K.~S.}\ \bibnamefont
  {Novoselov}}, \bibinfo {author} {\bibfnamefont {P.}~\bibnamefont {Dai}},
  \bibinfo {author} {\bibfnamefont {L.}~\bibnamefont {Balicas}},\ and\ \bibinfo
  {author} {\bibfnamefont {E.~J.}\ \bibnamefont {Santos}},\ }\bibfield  {title}
  {\bibinfo {title} {{The Magnetic Genome of Two-Dimensional van der Waals
  Materials}},\ }\href {https://doi.org/10.1021/ACSNANO.1C09150} {\bibfield
  {journal} {\bibinfo  {journal} {ACS Nano}\ }\textbf {\bibinfo {volume}
  {17}},\ \bibinfo {pages} {47} (\bibinfo {year} {2021})}\BibitemShut {NoStop}%
\bibitem [{\citenamefont {May}\ \emph {et~al.}(2019)\citenamefont {May},
  \citenamefont {Ovchinnikov}, \citenamefont {Zheng}, \citenamefont {Hermann},
  \citenamefont {Calder}, \citenamefont {Huang}, \citenamefont {Fei},
  \citenamefont {Liu}, \citenamefont {Xu},\ and\ \citenamefont
  {Mcguire}}]{May2019FerromagnetismArticle}%
  \BibitemOpen
  \bibfield  {author} {\bibinfo {author} {\bibfnamefont {A.~F.}\ \bibnamefont
  {May}}, \bibinfo {author} {\bibfnamefont {D.}~\bibnamefont {Ovchinnikov}},
  \bibinfo {author} {\bibfnamefont {Q.}~\bibnamefont {Zheng}}, \bibinfo
  {author} {\bibfnamefont {R.}~\bibnamefont {Hermann}}, \bibinfo {author}
  {\bibfnamefont {S.}~\bibnamefont {Calder}}, \bibinfo {author} {\bibfnamefont
  {B.}~\bibnamefont {Huang}}, \bibinfo {author} {\bibfnamefont
  {Z.}~\bibnamefont {Fei}}, \bibinfo {author} {\bibfnamefont {Y.}~\bibnamefont
  {Liu}}, \bibinfo {author} {\bibfnamefont {X.}~\bibnamefont {Xu}},\ and\
  \bibinfo {author} {\bibfnamefont {M.~A.}\ \bibnamefont {Mcguire}},\
  }\bibfield  {title} {\bibinfo {title} {Ferromagnetism near room temperature
  in the cleavable van der waals crystal {Fe}$_5${GeTe}$_2$},\ }\href
  {https://doi.org/10.1021/acsnano.8b09660} {\bibfield  {journal} {\bibinfo
  {journal} {ACS Nano}\ }\textbf {\bibinfo {volume} {13}},\ \bibinfo {pages}
  {57} (\bibinfo {year} {2019})}\BibitemShut {NoStop}%
\bibitem [{\citenamefont {Zhang}\ \emph {et~al.}(2020)\citenamefont {Zhang},
  \citenamefont {Chen}, \citenamefont {Zhai}, \citenamefont {Chen},
  \citenamefont {Caretta}, \citenamefont {Huang}, \citenamefont {Chopdekar},
  \citenamefont {Cao}, \citenamefont {Sun}, \citenamefont {Yao}, \citenamefont
  {Birgeneau},\ and\ \citenamefont {Ramesh}}]{Zhang2020ItinerantTemperature}%
  \BibitemOpen
  \bibfield  {author} {\bibinfo {author} {\bibfnamefont {H.}~\bibnamefont
  {Zhang}}, \bibinfo {author} {\bibfnamefont {R.}~\bibnamefont {Chen}},
  \bibinfo {author} {\bibfnamefont {K.}~\bibnamefont {Zhai}}, \bibinfo {author}
  {\bibfnamefont {X.}~\bibnamefont {Chen}}, \bibinfo {author} {\bibfnamefont
  {L.}~\bibnamefont {Caretta}}, \bibinfo {author} {\bibfnamefont
  {X.}~\bibnamefont {Huang}}, \bibinfo {author} {\bibfnamefont {R.~V.}\
  \bibnamefont {Chopdekar}}, \bibinfo {author} {\bibfnamefont {J.}~\bibnamefont
  {Cao}}, \bibinfo {author} {\bibfnamefont {J.}~\bibnamefont {Sun}}, \bibinfo
  {author} {\bibfnamefont {J.}~\bibnamefont {Yao}}, \bibinfo {author}
  {\bibfnamefont {R.}~\bibnamefont {Birgeneau}},\ and\ \bibinfo {author}
  {\bibfnamefont {R.}~\bibnamefont {Ramesh}},\ }\bibfield  {title} {\bibinfo
  {title} {{Itinerant ferromagnetism in van der Waals Fe$_{5-x}$GeTe$_2$
  crystals above room temperature}},\ }\href
  {https://doi.org/10.1103/PhysRevB.102.064417} {\bibfield  {journal} {\bibinfo
   {journal} {Phys. Rev. B}\ }\textbf {\bibinfo {volume} {102}},\ \bibinfo
  {pages} {64417} (\bibinfo {year} {2020})}\BibitemShut {NoStop}%
\bibitem [{\citenamefont {Chen}\ \emph {et~al.}(2023)\citenamefont {Chen},
  \citenamefont {Asif}, \citenamefont {Dolui}, \citenamefont {Wang},
  \citenamefont {T{\'a}mara-Isaza}, \citenamefont {Goli}, \citenamefont
  {Whalen}, \citenamefont {Wang}, \citenamefont {Chen}, \citenamefont {Zhang},
  \citenamefont {Liu}, \citenamefont {Jariwala}, \citenamefont {Jungfleisch},
  \citenamefont {Chakraborty}, \citenamefont {May}, \citenamefont {McGuire},
  \citenamefont {Nikolic}, \citenamefont {Xiao},\ and\ \citenamefont
  {Ku}}]{Chen2023Above-Room-Temperature2}%
  \BibitemOpen
  \bibfield  {author} {\bibinfo {author} {\bibfnamefont {H.}~\bibnamefont
  {Chen}}, \bibinfo {author} {\bibfnamefont {S.}~\bibnamefont {Asif}}, \bibinfo
  {author} {\bibfnamefont {K.}~\bibnamefont {Dolui}}, \bibinfo {author}
  {\bibfnamefont {Y.}~\bibnamefont {Wang}}, \bibinfo {author} {\bibfnamefont
  {J.}~\bibnamefont {T{\'a}mara-Isaza}}, \bibinfo {author} {\bibfnamefont
  {V.~M. L. D.~P.}\ \bibnamefont {Goli}}, \bibinfo {author} {\bibfnamefont
  {M.}~\bibnamefont {Whalen}}, \bibinfo {author} {\bibfnamefont
  {X.}~\bibnamefont {Wang}}, \bibinfo {author} {\bibfnamefont {Z.}~\bibnamefont
  {Chen}}, \bibinfo {author} {\bibfnamefont {H.}~\bibnamefont {Zhang}},
  \bibinfo {author} {\bibfnamefont {K.}~\bibnamefont {Liu}}, \bibinfo {author}
  {\bibfnamefont {D.}~\bibnamefont {Jariwala}}, \bibinfo {author}
  {\bibfnamefont {M.~B.}\ \bibnamefont {Jungfleisch}}, \bibinfo {author}
  {\bibfnamefont {C.}~\bibnamefont {Chakraborty}}, \bibinfo {author}
  {\bibfnamefont {A.~F.}\ \bibnamefont {May}}, \bibinfo {author} {\bibfnamefont
  {M.~A.}\ \bibnamefont {McGuire}}, \bibinfo {author} {\bibfnamefont {B.~K.}\
  \bibnamefont {Nikolic}}, \bibinfo {author} {\bibfnamefont {J.~Q.}\
  \bibnamefont {Xiao}},\ and\ \bibinfo {author} {\bibfnamefont {M.~J.~H.}\
  \bibnamefont {Ku}},\ }\bibfield  {title} {\bibinfo {title}
  {Above-room-temperature ferromagnetism in thin van der waals flakes of
  cobalt-substituted {Fe}$_5${GeTe}$_2$},\ }\href
  {https://pubs.acs.org/doi/10.1021/acsami.2c18028} {\bibfield  {journal}
  {\bibinfo  {journal} {ACS Applied Materials \& Interfaces}\ }\textbf
  {\bibinfo {volume} {15}},\ \bibinfo {pages} {3287} (\bibinfo {year}
  {2023})}\BibitemShut {NoStop}%
\bibitem [{\citenamefont {Wang}\ \emph {et~al.}(2018)\citenamefont {Wang},
  \citenamefont {Sapkota}, \citenamefont {Taniguchi}, \citenamefont {Watanabe},
  \citenamefont {Mandrus},\ and\ \citenamefont
  {Morpurgo}}]{Wang2018TunnelingHeterostructures}%
  \BibitemOpen
  \bibfield  {author} {\bibinfo {author} {\bibfnamefont {Z.}~\bibnamefont
  {Wang}}, \bibinfo {author} {\bibfnamefont {D.}~\bibnamefont {Sapkota}},
  \bibinfo {author} {\bibfnamefont {T.}~\bibnamefont {Taniguchi}}, \bibinfo
  {author} {\bibfnamefont {K.}~\bibnamefont {Watanabe}}, \bibinfo {author}
  {\bibfnamefont {D.}~\bibnamefont {Mandrus}},\ and\ \bibinfo {author}
  {\bibfnamefont {A.~F.}\ \bibnamefont {Morpurgo}},\ }\bibfield  {title}
  {\bibinfo {title} {Tunneling spin valves based on {Fe}$_3${GeTe}$_2$/
  h{BN}/{Fe}$_3${GeTe}$_2$ van der waals heterostructures},\ }\href
  {https://doi.org/10.1021/acs.nanolett.8b01278} {\bibfield  {journal}
  {\bibinfo  {journal} {Nano Letters}\ }\textbf {\bibinfo {volume} {18}},\
  \bibinfo {pages} {4303} (\bibinfo {year} {2018})}\BibitemShut {NoStop}%
\bibitem [{\citenamefont {Georgopoulou-Kotsaki}\ \emph
  {et~al.}(2023)\citenamefont {Georgopoulou-Kotsaki}, \citenamefont {Pappas},
  \citenamefont {Lintzeris}, \citenamefont {Tsipas}, \citenamefont {Fragkos},
  \citenamefont {Markou}, \citenamefont {Felser}, \citenamefont {Longo},
  \citenamefont {Fanciulli}, \citenamefont {Mantovan}, \citenamefont
  {Mahfouzi}, \citenamefont {Kioussis},\ and\ \citenamefont
  {Dimoulas}}]{Georgopoulou-Kotsaki2023Significant}%
  \BibitemOpen
  \bibfield  {author} {\bibinfo {author} {\bibfnamefont {E.}~\bibnamefont
  {Georgopoulou-Kotsaki}}, \bibinfo {author} {\bibfnamefont {P.}~\bibnamefont
  {Pappas}}, \bibinfo {author} {\bibfnamefont {A.}~\bibnamefont {Lintzeris}},
  \bibinfo {author} {\bibfnamefont {P.}~\bibnamefont {Tsipas}}, \bibinfo
  {author} {\bibfnamefont {S.}~\bibnamefont {Fragkos}}, \bibinfo {author}
  {\bibfnamefont {A.}~\bibnamefont {Markou}}, \bibinfo {author} {\bibfnamefont
  {C.}~\bibnamefont {Felser}}, \bibinfo {author} {\bibfnamefont
  {E.}~\bibnamefont {Longo}}, \bibinfo {author} {\bibfnamefont
  {M.}~\bibnamefont {Fanciulli}}, \bibinfo {author} {\bibfnamefont
  {R.}~\bibnamefont {Mantovan}}, \bibinfo {author} {\bibfnamefont
  {F.}~\bibnamefont {Mahfouzi}}, \bibinfo {author} {\bibfnamefont
  {N.}~\bibnamefont {Kioussis}},\ and\ \bibinfo {author} {\bibfnamefont
  {A.}~\bibnamefont {Dimoulas}},\ }\bibfield  {title} {\bibinfo {title}
  {Significant enhancement of ferromagnetism above room temperature in
  epitaxial 2d van der waals ferromagnet
  {Fe}$_{5-\delta}${GeTe}$_2$/{Bi}$_2${Te}$_3$ heterostructures},\ }\href
  {https://doi.org/10.1039/D2NR04820E} {\bibfield  {journal} {\bibinfo
  {journal} {Nanoscale}\ }\textbf {\bibinfo {volume} {15}},\ \bibinfo {pages}
  {2223} (\bibinfo {year} {2023})}\BibitemShut {NoStop}%
\bibitem [{\citenamefont {Li}\ \emph {et~al.}(2018)\citenamefont {Li},
  \citenamefont {Yang}, \citenamefont {Gong}, \citenamefont {Chopdekar},
  \citenamefont {N'diaye}, \citenamefont {Turner}, \citenamefont {Chen},
  \citenamefont {Scholl}, \citenamefont {Shafer}, \citenamefont {Arenholz},
  \citenamefont {Schmid}, \citenamefont {Wang}, \citenamefont {Liu},
  \citenamefont {Gao}, \citenamefont {Admasu}, \citenamefont {Cheong},
  \citenamefont {Hwang}, \citenamefont {Li}, \citenamefont {Wang},
  \citenamefont {Zhang},\ and\ \citenamefont {Qiu}}]{Li2018Patterning}%
  \BibitemOpen
  \bibfield  {author} {\bibinfo {author} {\bibfnamefont {Q.}~\bibnamefont
  {Li}}, \bibinfo {author} {\bibfnamefont {M.}~\bibnamefont {Yang}}, \bibinfo
  {author} {\bibfnamefont {C.}~\bibnamefont {Gong}}, \bibinfo {author}
  {\bibfnamefont {R.~V.}\ \bibnamefont {Chopdekar}}, \bibinfo {author}
  {\bibfnamefont {A.~T.}\ \bibnamefont {N'diaye}}, \bibinfo {author}
  {\bibfnamefont {J.}~\bibnamefont {Turner}}, \bibinfo {author} {\bibfnamefont
  {G.}~\bibnamefont {Chen}}, \bibinfo {author} {\bibfnamefont {A.}~\bibnamefont
  {Scholl}}, \bibinfo {author} {\bibfnamefont {P.}~\bibnamefont {Shafer}},
  \bibinfo {author} {\bibfnamefont {E.}~\bibnamefont {Arenholz}}, \bibinfo
  {author} {\bibfnamefont {A.~K.}\ \bibnamefont {Schmid}}, \bibinfo {author}
  {\bibfnamefont {S.}~\bibnamefont {Wang}}, \bibinfo {author} {\bibfnamefont
  {K.}~\bibnamefont {Liu}}, \bibinfo {author} {\bibfnamefont {N.}~\bibnamefont
  {Gao}}, \bibinfo {author} {\bibfnamefont {A.~S.}\ \bibnamefont {Admasu}},
  \bibinfo {author} {\bibfnamefont {S.-W.}\ \bibnamefont {Cheong}}, \bibinfo
  {author} {\bibfnamefont {C.}~\bibnamefont {Hwang}}, \bibinfo {author}
  {\bibfnamefont {J.}~\bibnamefont {Li}}, \bibinfo {author} {\bibfnamefont
  {F.}~\bibnamefont {Wang}}, \bibinfo {author} {\bibfnamefont {X.}~\bibnamefont
  {Zhang}},\ and\ \bibinfo {author} {\bibfnamefont {Z.}~\bibnamefont {Qiu}},\
  }\bibfield  {title} {\bibinfo {title} {Patterning-induced ferromagnetism of
  {Fe}$_3${GeTe}$_2$ van der waals materials beyond room temperature},\ }\href
  {https://doi.org/10.1021/acs.nanolett.8b02806} {\bibfield  {journal}
  {\bibinfo  {journal} {Nano Lett}\ }\textbf {\bibinfo {volume} {18}},\
  \bibinfo {pages} {7} (\bibinfo {year} {2018})}\BibitemShut {NoStop}%
\bibitem [{\citenamefont {Alghamdi}\ \emph {et~al.}(2019)\citenamefont
  {Alghamdi}, \citenamefont {Lohmann}, \citenamefont {Li}, \citenamefont
  {Jothi}, \citenamefont {Shao}, \citenamefont {Aldosary}, \citenamefont {Su},
  \citenamefont {Fokwa},\ and\ \citenamefont {Shi}}]{Alghamdi2019Highly}%
  \BibitemOpen
  \bibfield  {author} {\bibinfo {author} {\bibfnamefont {M.}~\bibnamefont
  {Alghamdi}}, \bibinfo {author} {\bibfnamefont {M.}~\bibnamefont {Lohmann}},
  \bibinfo {author} {\bibfnamefont {J.}~\bibnamefont {Li}}, \bibinfo {author}
  {\bibfnamefont {P.~R.}\ \bibnamefont {Jothi}}, \bibinfo {author}
  {\bibfnamefont {Q.}~\bibnamefont {Shao}}, \bibinfo {author} {\bibfnamefont
  {M.}~\bibnamefont {Aldosary}}, \bibinfo {author} {\bibfnamefont
  {T.}~\bibnamefont {Su}}, \bibinfo {author} {\bibfnamefont {B.~P.~T.}\
  \bibnamefont {Fokwa}},\ and\ \bibinfo {author} {\bibfnamefont
  {J.}~\bibnamefont {Shi}},\ }\bibfield  {title} {\bibinfo {title} {Highly
  efficient spin-orbit torque and switching of layered ferromagnet
  {Fe}$_3${GeTe}$_2$},\ }\href {https://doi.org/10.1021/acs.nanolett.9b01043}
  {\bibfield  {journal} {\bibinfo  {journal} {Nano Lett}\ }\textbf {\bibinfo
  {volume} {19}},\ \bibinfo {pages} {25} (\bibinfo {year} {2019})}\BibitemShut
  {NoStop}%
\bibitem [{\citenamefont {Yang}\ \emph {et~al.}(2020)\citenamefont {Yang},
  \citenamefont {Li}, \citenamefont {Chopdekar}, \citenamefont {Stan},
  \citenamefont {Cabrini}, \citenamefont {Choi}, \citenamefont {Wang},
  \citenamefont {Wang}, \citenamefont {Gao}, \citenamefont {Scholl},
  \citenamefont {Tamura}, \citenamefont {Hwang}, \citenamefont {Wang},\ and\
  \citenamefont {Qiu}}]{Yang2020Highly}%
  \BibitemOpen
  \bibfield  {author} {\bibinfo {author} {\bibfnamefont {M.}~\bibnamefont
  {Yang}}, \bibinfo {author} {\bibfnamefont {Q.}~\bibnamefont {Li}}, \bibinfo
  {author} {\bibfnamefont {R.~V.}\ \bibnamefont {Chopdekar}}, \bibinfo {author}
  {\bibfnamefont {C.}~\bibnamefont {Stan}}, \bibinfo {author} {\bibfnamefont
  {S.}~\bibnamefont {Cabrini}}, \bibinfo {author} {\bibfnamefont {J.~W.}\
  \bibnamefont {Choi}}, \bibinfo {author} {\bibfnamefont {S.}~\bibnamefont
  {Wang}}, \bibinfo {author} {\bibfnamefont {T.}~\bibnamefont {Wang}}, \bibinfo
  {author} {\bibfnamefont {N.}~\bibnamefont {Gao}}, \bibinfo {author}
  {\bibfnamefont {A.}~\bibnamefont {Scholl}}, \bibinfo {author} {\bibfnamefont
  {N.}~\bibnamefont {Tamura}}, \bibinfo {author} {\bibfnamefont
  {C.}~\bibnamefont {Hwang}}, \bibinfo {author} {\bibfnamefont
  {F.}~\bibnamefont {Wang}},\ and\ \bibinfo {author} {\bibfnamefont
  {Z.}~\bibnamefont {Qiu}},\ }\bibfield  {title} {\bibinfo {title} {Highly
  enhanced curie temperature in ga-implanted {Fe}$_3${GeTe}$_2$ van der waals
  material},\ }\href
  {https://onlinelibrary.wiley.com/doi/full/10.1002/qute.202000017} {\bibfield
  {journal} {\bibinfo  {journal} {Advanced Quantum Technologies}\ }\textbf
  {\bibinfo {volume} {3}} (\bibinfo {year} {2020})}\BibitemShut {NoStop}%
\bibitem [{\citenamefont {Ding}\ \emph {et~al.}(2020)\citenamefont {Ding},
  \citenamefont {Li}, \citenamefont {Xu}, \citenamefont {Li}, \citenamefont
  {Hou}, \citenamefont {Liu}, \citenamefont {Xi}, \citenamefont {Xu},
  \citenamefont {Yao},\ and\ \citenamefont {Wang}}]{Ding2020Observtion}%
  \BibitemOpen
  \bibfield  {author} {\bibinfo {author} {\bibfnamefont {B.}~\bibnamefont
  {Ding}}, \bibinfo {author} {\bibfnamefont {Z.}~\bibnamefont {Li}}, \bibinfo
  {author} {\bibfnamefont {G.}~\bibnamefont {Xu}}, \bibinfo {author}
  {\bibfnamefont {H.}~\bibnamefont {Li}}, \bibinfo {author} {\bibfnamefont
  {Z.}~\bibnamefont {Hou}}, \bibinfo {author} {\bibfnamefont {E.}~\bibnamefont
  {Liu}}, \bibinfo {author} {\bibfnamefont {X.}~\bibnamefont {Xi}}, \bibinfo
  {author} {\bibfnamefont {F.}~\bibnamefont {Xu}}, \bibinfo {author}
  {\bibfnamefont {Y.}~\bibnamefont {Yao}},\ and\ \bibinfo {author}
  {\bibfnamefont {W.}~\bibnamefont {Wang}},\ }\bibfield  {title} {\bibinfo
  {title} {Observation of magnetic skyrmion bubbles in a van der waals
  ferromagnet {Fe}$_3${GeTe}$_2$},\ }\href
  {https://doi.org/10.1021/acs.nanolett.9b03453} {\bibfield  {journal}
  {\bibinfo  {journal} {Nano Lett}\ }\textbf {\bibinfo {volume} {20}},\
  \bibinfo {pages} {2023} (\bibinfo {year} {2020})}\BibitemShut {NoStop}%
\bibitem [{\citenamefont {Wu}\ \emph {et~al.}(2020)\citenamefont {Wu},
  \citenamefont {Zhang}, \citenamefont {Zhang}, \citenamefont {Wang},
  \citenamefont {Zhu}, \citenamefont {Hu}, \citenamefont {Yin}, \citenamefont
  {Wong}, \citenamefont {Fang}, \citenamefont {Wan}, \citenamefont {Han},
  \citenamefont {Shao}, \citenamefont {Taniguchi}, \citenamefont {Watanabe},
  \citenamefont {Zang}, \citenamefont {Mao}, \citenamefont {Zhang},\ and\
  \citenamefont {Wang}}]{Wu2020Neel}%
  \BibitemOpen
  \bibfield  {author} {\bibinfo {author} {\bibfnamefont {Y.}~\bibnamefont
  {Wu}}, \bibinfo {author} {\bibfnamefont {S.}~\bibnamefont {Zhang}}, \bibinfo
  {author} {\bibfnamefont {J.}~\bibnamefont {Zhang}}, \bibinfo {author}
  {\bibfnamefont {W.}~\bibnamefont {Wang}}, \bibinfo {author} {\bibfnamefont
  {Y.~L.}\ \bibnamefont {Zhu}}, \bibinfo {author} {\bibfnamefont
  {J.}~\bibnamefont {Hu}}, \bibinfo {author} {\bibfnamefont {G.}~\bibnamefont
  {Yin}}, \bibinfo {author} {\bibfnamefont {K.}~\bibnamefont {Wong}}, \bibinfo
  {author} {\bibfnamefont {C.}~\bibnamefont {Fang}}, \bibinfo {author}
  {\bibfnamefont {C.}~\bibnamefont {Wan}}, \bibinfo {author} {\bibfnamefont
  {X.}~\bibnamefont {Han}}, \bibinfo {author} {\bibfnamefont {Q.}~\bibnamefont
  {Shao}}, \bibinfo {author} {\bibfnamefont {T.}~\bibnamefont {Taniguchi}},
  \bibinfo {author} {\bibfnamefont {K.}~\bibnamefont {Watanabe}}, \bibinfo
  {author} {\bibfnamefont {J.}~\bibnamefont {Zang}}, \bibinfo {author}
  {\bibfnamefont {Z.}~\bibnamefont {Mao}}, \bibinfo {author} {\bibfnamefont
  {X.}~\bibnamefont {Zhang}},\ and\ \bibinfo {author} {\bibfnamefont {K.~L.}\
  \bibnamefont {Wang}},\ }\bibfield  {title} {\bibinfo {title}
  {{N{\'{e}}el-type skyrmion in WTe$_2$/Fe$_3$GeTe$_2$ van der Waals
  heterostructure}},\ }\href {https://doi.org/10.1038/s41467-020-17566-x}
  {\bibfield  {journal} {\bibinfo  {journal} {Nature Communications 2020 11:1}\
  }\textbf {\bibinfo {volume} {11}},\ \bibinfo {pages} {1} (\bibinfo {year}
  {2020})}\BibitemShut {NoStop}%
\bibitem [{\citenamefont {Deng}\ \emph {et~al.}(2018)\citenamefont {Deng},
  \citenamefont {Yu}, \citenamefont {Song}, \citenamefont {Zhang},
  \citenamefont {Wang}, \citenamefont {Sun}, \citenamefont {Yi}, \citenamefont
  {Wu}, \citenamefont {Wu}, \citenamefont {Zhu}, \citenamefont {Wang},
  \citenamefont {Chen},\ and\ \citenamefont
  {Zhang}}]{Deng2018Gate-tunableFe3GeTe2}%
  \BibitemOpen
  \bibfield  {author} {\bibinfo {author} {\bibfnamefont {Y.}~\bibnamefont
  {Deng}}, \bibinfo {author} {\bibfnamefont {Y.}~\bibnamefont {Yu}}, \bibinfo
  {author} {\bibfnamefont {Y.}~\bibnamefont {Song}}, \bibinfo {author}
  {\bibfnamefont {J.}~\bibnamefont {Zhang}}, \bibinfo {author} {\bibfnamefont
  {N.~Z.}\ \bibnamefont {Wang}}, \bibinfo {author} {\bibfnamefont
  {Z.}~\bibnamefont {Sun}}, \bibinfo {author} {\bibfnamefont {Y.}~\bibnamefont
  {Yi}}, \bibinfo {author} {\bibfnamefont {Y.~Z.}\ \bibnamefont {Wu}}, \bibinfo
  {author} {\bibfnamefont {S.}~\bibnamefont {Wu}}, \bibinfo {author}
  {\bibfnamefont {J.}~\bibnamefont {Zhu}}, \bibinfo {author} {\bibfnamefont
  {J.}~\bibnamefont {Wang}}, \bibinfo {author} {\bibfnamefont {X.~H.}\
  \bibnamefont {Chen}},\ and\ \bibinfo {author} {\bibfnamefont
  {Y.}~\bibnamefont {Zhang}},\ }\bibfield  {title} {\bibinfo {title}
  {{Gate-tunable room-temperature ferromagnetism in two-dimensional
  Fe$_3$GeTe$_2$}},\ }\href {https://doi.org/10.1038/s41586-018-0626-9}
  {\bibfield  {journal} {\bibinfo  {journal} {Nature}\ }\textbf {\bibinfo
  {volume} {563}},\ \bibinfo {pages} {94} (\bibinfo {year} {2018})}\BibitemShut
  {NoStop}%
\bibitem [{\citenamefont {Zhang}\ \emph {et~al.}(2018)\citenamefont {Zhang},
  \citenamefont {Lu}, \citenamefont {Zhu}, \citenamefont {Tan}, \citenamefont
  {Feng}, \citenamefont {Liu}, \citenamefont {Zhang}, \citenamefont {Chen},
  \citenamefont {Liu}, \citenamefont {Luo}, \citenamefont {Xie}, \citenamefont
  {Luo}, \citenamefont {Zhang},\ and\ \citenamefont {Lai}}]{Yun2018Emergence}%
  \BibitemOpen
  \bibfield  {author} {\bibinfo {author} {\bibfnamefont {Y.}~\bibnamefont
  {Zhang}}, \bibinfo {author} {\bibfnamefont {H.}~\bibnamefont {Lu}}, \bibinfo
  {author} {\bibfnamefont {X.}~\bibnamefont {Zhu}}, \bibinfo {author}
  {\bibfnamefont {S.}~\bibnamefont {Tan}}, \bibinfo {author} {\bibfnamefont
  {W.}~\bibnamefont {Feng}}, \bibinfo {author} {\bibfnamefont {Q.}~\bibnamefont
  {Liu}}, \bibinfo {author} {\bibfnamefont {W.}~\bibnamefont {Zhang}}, \bibinfo
  {author} {\bibfnamefont {Q.}~\bibnamefont {Chen}}, \bibinfo {author}
  {\bibfnamefont {Y.}~\bibnamefont {Liu}}, \bibinfo {author} {\bibfnamefont
  {X.}~\bibnamefont {Luo}}, \bibinfo {author} {\bibfnamefont {D.}~\bibnamefont
  {Xie}}, \bibinfo {author} {\bibfnamefont {L.}~\bibnamefont {Luo}}, \bibinfo
  {author} {\bibfnamefont {Z.}~\bibnamefont {Zhang}},\ and\ \bibinfo {author}
  {\bibfnamefont {X.}~\bibnamefont {Lai}},\ }\bibfield  {title} {\bibinfo
  {title} {Emergence of kondo lattice behavior in a van der waals itinerant
  ferromagnet, {Fe}$_3${GeTe}$_2$},\ }\href
  {https://doi.org/10.1126/sciadv.aao6791} {\bibfield  {journal} {\bibinfo
  {journal} {Science Advances}\ }\textbf {\bibinfo {volume} {4}},\ \bibinfo
  {pages} {eaao6791} (\bibinfo {year} {2018})}\BibitemShut {NoStop}%
\bibitem [{\citenamefont {Bao}\ \emph {et~al.}(2022)\citenamefont {Bao},
  \citenamefont {Wang}, \citenamefont {Shangguan}, \citenamefont {Cai},
  \citenamefont {Dong}, \citenamefont {Huang}, \citenamefont {Si},
  \citenamefont {Ma}, \citenamefont {Kajimoto}, \citenamefont {Ikeuchi},
  \citenamefont {Yano}, \citenamefont {Yu}, \citenamefont {Wan}, \citenamefont
  {Li},\ and\ \citenamefont {Wen}}]{Bao2022Neutron2}%
  \BibitemOpen
  \bibfield  {author} {\bibinfo {author} {\bibfnamefont {S.}~\bibnamefont
  {Bao}}, \bibinfo {author} {\bibfnamefont {W.}~\bibnamefont {Wang}}, \bibinfo
  {author} {\bibfnamefont {Y.}~\bibnamefont {Shangguan}}, \bibinfo {author}
  {\bibfnamefont {Z.}~\bibnamefont {Cai}}, \bibinfo {author} {\bibfnamefont
  {Z.-Y.}\ \bibnamefont {Dong}}, \bibinfo {author} {\bibfnamefont
  {Z.}~\bibnamefont {Huang}}, \bibinfo {author} {\bibfnamefont
  {W.}~\bibnamefont {Si}}, \bibinfo {author} {\bibfnamefont {Z.}~\bibnamefont
  {Ma}}, \bibinfo {author} {\bibfnamefont {R.}~\bibnamefont {Kajimoto}},
  \bibinfo {author} {\bibfnamefont {K.}~\bibnamefont {Ikeuchi}}, \bibinfo
  {author} {\bibfnamefont {S.-i.}\ \bibnamefont {Yano}}, \bibinfo {author}
  {\bibfnamefont {S.-L.}\ \bibnamefont {Yu}}, \bibinfo {author} {\bibfnamefont
  {X.}~\bibnamefont {Wan}}, \bibinfo {author} {\bibfnamefont {J.-X.}\
  \bibnamefont {Li}},\ and\ \bibinfo {author} {\bibfnamefont {J.}~\bibnamefont
  {Wen}},\ }\bibfield  {title} {\bibinfo {title} {Neutron spectroscopy evidence
  on the dual nature of magnetic excitations in a van der waals metallic
  ferromagnet {Fe}$_{2.72}${GeTe}$_2$},\ }\href
  {https://doi.org/10.1103/PhysRevX.12.011022} {\bibfield  {journal} {\bibinfo
  {journal} {Phys. Rev. X}\ }\textbf {\bibinfo {volume} {12}},\ \bibinfo
  {pages} {011022} (\bibinfo {year} {2022})}\BibitemShut {NoStop}%
\bibitem [{\citenamefont {Wu}\ \emph {et~al.}(2021)\citenamefont {Wu},
  \citenamefont {Lei}, \citenamefont {Yin}, \citenamefont {Zhao}, \citenamefont
  {Li}, \citenamefont {Wang}, \citenamefont {Liu}, \citenamefont {Song},
  \citenamefont {Ma}, \citenamefont {Ding}, \citenamefont {Cheng},
  \citenamefont {Liu}, \citenamefont {Lei},\ and\ \citenamefont
  {Wang}}]{Wu2021DirectGeTe_2}%
  \BibitemOpen
  \bibfield  {author} {\bibinfo {author} {\bibfnamefont {X.}~\bibnamefont
  {Wu}}, \bibinfo {author} {\bibfnamefont {L.}~\bibnamefont {Lei}}, \bibinfo
  {author} {\bibfnamefont {Q.}~\bibnamefont {Yin}}, \bibinfo {author}
  {\bibfnamefont {N.-N.}\ \bibnamefont {Zhao}}, \bibinfo {author}
  {\bibfnamefont {M.}~\bibnamefont {Li}}, \bibinfo {author} {\bibfnamefont
  {Z.}~\bibnamefont {Wang}}, \bibinfo {author} {\bibfnamefont {Q.}~\bibnamefont
  {Liu}}, \bibinfo {author} {\bibfnamefont {W.}~\bibnamefont {Song}}, \bibinfo
  {author} {\bibfnamefont {H.}~\bibnamefont {Ma}}, \bibinfo {author}
  {\bibfnamefont {P.}~\bibnamefont {Ding}}, \bibinfo {author} {\bibfnamefont
  {Z.}~\bibnamefont {Cheng}}, \bibinfo {author} {\bibfnamefont
  {K.}~\bibnamefont {Liu}}, \bibinfo {author} {\bibfnamefont {H.}~\bibnamefont
  {Lei}},\ and\ \bibinfo {author} {\bibfnamefont {S.}~\bibnamefont {Wang}},\
  }\bibfield  {title} {\bibinfo {title} {Direct observation of competition
  between charge order and itinerant ferromagnetism in the van der waals
  crystal {Fe}$_{5-x}${GeTe}$_2$},\ }\href
  {https://doi.org/10.1103/PhysRevB.104.165101} {\bibfield  {journal} {\bibinfo
   {journal} {Phys. Rev. B}\ }\textbf {\bibinfo {volume} {104}},\ \bibinfo
  {pages} {165101} (\bibinfo {year} {2021})}\BibitemShut {NoStop}%
\bibitem [{\citenamefont {Wu}\ \emph {et~al.}(2024)\citenamefont {Wu},
  \citenamefont {Chen}, \citenamefont {Malinowski}, \citenamefont {Jang},
  \citenamefont {Deng}, \citenamefont {Scott}, \citenamefont {Huang},
  \citenamefont {Ruff}, \citenamefont {He}, \citenamefont {Chen}, \citenamefont
  {Hu}, \citenamefont {Yue}, \citenamefont {Oh}, \citenamefont {Teng},
  \citenamefont {Guo}, \citenamefont {Klemm}, \citenamefont {Shi},
  \citenamefont {Shi}, \citenamefont {Setty}, \citenamefont {Werner},
  \citenamefont {Hashimoto}, \citenamefont {Lu}, \citenamefont {Yilmaz},
  \citenamefont {Vescovo}, \citenamefont {Mo}, \citenamefont {Fedorov},
  \citenamefont {Denlinger}, \citenamefont {Xie}, \citenamefont {Gao},
  \citenamefont {Kono}, \citenamefont {Dai}, \citenamefont {Han}, \citenamefont
  {Xu}, \citenamefont {Birgeneau}, \citenamefont {Zhu}, \citenamefont
  {da~Silva~Neto}, \citenamefont {Wu}, \citenamefont {Chu}, \citenamefont
  {Si},\ and\ \citenamefont {Yi}}]{Wu2023ReversibleFerromagnet}%
  \BibitemOpen
  \bibfield  {author} {\bibinfo {author} {\bibfnamefont {H.}~\bibnamefont
  {Wu}}, \bibinfo {author} {\bibfnamefont {L.}~\bibnamefont {Chen}}, \bibinfo
  {author} {\bibfnamefont {P.}~\bibnamefont {Malinowski}}, \bibinfo {author}
  {\bibfnamefont {B.~G.}\ \bibnamefont {Jang}}, \bibinfo {author}
  {\bibfnamefont {Q.}~\bibnamefont {Deng}}, \bibinfo {author} {\bibfnamefont
  {K.}~\bibnamefont {Scott}}, \bibinfo {author} {\bibfnamefont
  {J.}~\bibnamefont {Huang}}, \bibinfo {author} {\bibfnamefont {J.~P.~C.}\
  \bibnamefont {Ruff}}, \bibinfo {author} {\bibfnamefont {Y.}~\bibnamefont
  {He}}, \bibinfo {author} {\bibfnamefont {X.}~\bibnamefont {Chen}}, \bibinfo
  {author} {\bibfnamefont {C.}~\bibnamefont {Hu}}, \bibinfo {author}
  {\bibfnamefont {Z.}~\bibnamefont {Yue}}, \bibinfo {author} {\bibfnamefont
  {J.~S.}\ \bibnamefont {Oh}}, \bibinfo {author} {\bibfnamefont
  {X.}~\bibnamefont {Teng}}, \bibinfo {author} {\bibfnamefont {Y.}~\bibnamefont
  {Guo}}, \bibinfo {author} {\bibfnamefont {M.}~\bibnamefont {Klemm}}, \bibinfo
  {author} {\bibfnamefont {C.}~\bibnamefont {Shi}}, \bibinfo {author}
  {\bibfnamefont {Y.}~\bibnamefont {Shi}}, \bibinfo {author} {\bibfnamefont
  {C.}~\bibnamefont {Setty}}, \bibinfo {author} {\bibfnamefont
  {T.}~\bibnamefont {Werner}}, \bibinfo {author} {\bibfnamefont
  {M.}~\bibnamefont {Hashimoto}}, \bibinfo {author} {\bibfnamefont
  {D.}~\bibnamefont {Lu}}, \bibinfo {author} {\bibfnamefont {T.}~\bibnamefont
  {Yilmaz}}, \bibinfo {author} {\bibfnamefont {E.}~\bibnamefont {Vescovo}},
  \bibinfo {author} {\bibfnamefont {S.-K.}\ \bibnamefont {Mo}}, \bibinfo
  {author} {\bibfnamefont {A.}~\bibnamefont {Fedorov}}, \bibinfo {author}
  {\bibfnamefont {J.~D.}\ \bibnamefont {Denlinger}}, \bibinfo {author}
  {\bibfnamefont {Y.}~\bibnamefont {Xie}}, \bibinfo {author} {\bibfnamefont
  {B.}~\bibnamefont {Gao}}, \bibinfo {author} {\bibfnamefont {J.}~\bibnamefont
  {Kono}}, \bibinfo {author} {\bibfnamefont {P.}~\bibnamefont {Dai}}, \bibinfo
  {author} {\bibfnamefont {Y.}~\bibnamefont {Han}}, \bibinfo {author}
  {\bibfnamefont {X.}~\bibnamefont {Xu}}, \bibinfo {author} {\bibfnamefont
  {R.~J.}\ \bibnamefont {Birgeneau}}, \bibinfo {author} {\bibfnamefont {J.-X.}\
  \bibnamefont {Zhu}}, \bibinfo {author} {\bibfnamefont {E.~H.}\ \bibnamefont
  {da~Silva~Neto}}, \bibinfo {author} {\bibfnamefont {L.}~\bibnamefont {Wu}},
  \bibinfo {author} {\bibfnamefont {J.-H.}\ \bibnamefont {Chu}}, \bibinfo
  {author} {\bibfnamefont {Q.}~\bibnamefont {Si}},\ and\ \bibinfo {author}
  {\bibfnamefont {M.}~\bibnamefont {Yi}},\ }\bibfield  {title} {\bibinfo
  {title} {Reversible non-volatile electronic switching in a
  near-room-temperature van der waals ferromagnet},\ }\href
  {https://doi.org/10.1038/s41467-024-46862-z} {\bibfield  {journal} {\bibinfo
  {journal} {Nat. Comm.}\ }\textbf {\bibinfo {volume} {15}},\ \bibinfo {pages}
  {2739} (\bibinfo {year} {2024})}\BibitemShut {NoStop}%
\bibitem [{\citenamefont {Kato}\ \emph {et~al.}(2022)\citenamefont {Kato},
  \citenamefont {Okamura}, \citenamefont {Minami}, \citenamefont {Fujimura},
  \citenamefont {Mogi}, \citenamefont {Yoshimi}, \citenamefont {Tsukazaki},
  \citenamefont {Takahashi}, \citenamefont {Kawasaki}, \citenamefont {Arita},
  \citenamefont {Tokura},\ and\ \citenamefont {Takahashi}}]{Kato_npj2022}%
  \BibitemOpen
  \bibfield  {author} {\bibinfo {author} {\bibfnamefont {Y.~D.}\ \bibnamefont
  {Kato}}, \bibinfo {author} {\bibfnamefont {Y.}~\bibnamefont {Okamura}},
  \bibinfo {author} {\bibfnamefont {S.}~\bibnamefont {Minami}}, \bibinfo
  {author} {\bibfnamefont {R.}~\bibnamefont {Fujimura}}, \bibinfo {author}
  {\bibfnamefont {M.}~\bibnamefont {Mogi}}, \bibinfo {author} {\bibfnamefont
  {R.}~\bibnamefont {Yoshimi}}, \bibinfo {author} {\bibfnamefont
  {A.}~\bibnamefont {Tsukazaki}}, \bibinfo {author} {\bibfnamefont {K.~S.}\
  \bibnamefont {Takahashi}}, \bibinfo {author} {\bibfnamefont {M.}~\bibnamefont
  {Kawasaki}}, \bibinfo {author} {\bibfnamefont {R.}~\bibnamefont {Arita}},
  \bibinfo {author} {\bibfnamefont {Y.}~\bibnamefont {Tokura}},\ and\ \bibinfo
  {author} {\bibfnamefont {Y.}~\bibnamefont {Takahashi}},\ }\bibfield  {title}
  {\bibinfo {title} {Optical anomalous hall effect enhanced by flat bands in
  ferromagnetic van der waals semimetal},\ }\href
  {https://www.nature.com/articles/s41535-022-00482-2} {\bibfield  {journal}
  {\bibinfo  {journal} {npj Quantum Materials}\ }\textbf {\bibinfo {volume}
  {7}},\ \bibinfo {pages} {73} (\bibinfo {year} {2022})}\BibitemShut {NoStop}%
\bibitem [{\citenamefont {Bai}\ \emph {et~al.}(2022)\citenamefont {Bai},
  \citenamefont {Lechermann}, \citenamefont {Liu}, \citenamefont {Cheng},
  \citenamefont {Kolesnikov}, \citenamefont {Ye}, \citenamefont {Williams},
  \citenamefont {Chi}, \citenamefont {Hong}, \citenamefont {Granroth},
  \citenamefont {May},\ and\ \citenamefont
  {Calder}}]{Bai2022AntiferromagnetictextGeTe_2}%
  \BibitemOpen
  \bibfield  {author} {\bibinfo {author} {\bibfnamefont {X.}~\bibnamefont
  {Bai}}, \bibinfo {author} {\bibfnamefont {F.}~\bibnamefont {Lechermann}},
  \bibinfo {author} {\bibfnamefont {Y.}~\bibnamefont {Liu}}, \bibinfo {author}
  {\bibfnamefont {Y.}~\bibnamefont {Cheng}}, \bibinfo {author} {\bibfnamefont
  {A.~I.}\ \bibnamefont {Kolesnikov}}, \bibinfo {author} {\bibfnamefont
  {F.}~\bibnamefont {Ye}}, \bibinfo {author} {\bibfnamefont {T.~J.}\
  \bibnamefont {Williams}}, \bibinfo {author} {\bibfnamefont {S.}~\bibnamefont
  {Chi}}, \bibinfo {author} {\bibfnamefont {T.}~\bibnamefont {Hong}}, \bibinfo
  {author} {\bibfnamefont {G.~E.}\ \bibnamefont {Granroth}}, \bibinfo {author}
  {\bibfnamefont {A.~F.}\ \bibnamefont {May}},\ and\ \bibinfo {author}
  {\bibfnamefont {S.}~\bibnamefont {Calder}},\ }\bibfield  {title} {\bibinfo
  {title} {Antiferromagnetic fluctuations and orbital-selective mott transition
  in the van der waals ferromagnet {Fe}$_{3-x}${GeTe}$_2$},\ }\href
  {https://doi.org/10.1103/PhysRevB.106.L180409} {\bibfield  {journal}
  {\bibinfo  {journal} {Phys. Rev. B}\ }\textbf {\bibinfo {volume} {106}},\
  \bibinfo {pages} {L180409} (\bibinfo {year} {2022})}\BibitemShut {NoStop}%
\bibitem [{\citenamefont {Xu}\ \emph {et~al.}(2020)\citenamefont {Xu},
  \citenamefont {Li}, \citenamefont {Duan}, \citenamefont {Zhang},
  \citenamefont {Chen}, \citenamefont {Kang}, \citenamefont {Liang},
  \citenamefont {Chen}, \citenamefont {Xia}, \citenamefont {Xu}, \citenamefont
  {Malinowski}, \citenamefont {Xu}, \citenamefont {Chu}, \citenamefont {Li},
  \citenamefont {Guo}, \citenamefont {Liu}, \citenamefont {Yang},\ and\
  \citenamefont {Chen}}]{Xu2020}%
  \BibitemOpen
  \bibfield  {author} {\bibinfo {author} {\bibfnamefont {X.}~\bibnamefont
  {Xu}}, \bibinfo {author} {\bibfnamefont {Y.~W.}\ \bibnamefont {Li}}, \bibinfo
  {author} {\bibfnamefont {S.~R.}\ \bibnamefont {Duan}}, \bibinfo {author}
  {\bibfnamefont {S.~L.}\ \bibnamefont {Zhang}}, \bibinfo {author}
  {\bibfnamefont {Y.~J.}\ \bibnamefont {Chen}}, \bibinfo {author}
  {\bibfnamefont {L.}~\bibnamefont {Kang}}, \bibinfo {author} {\bibfnamefont
  {A.~J.}\ \bibnamefont {Liang}}, \bibinfo {author} {\bibfnamefont
  {C.}~\bibnamefont {Chen}}, \bibinfo {author} {\bibfnamefont {W.}~\bibnamefont
  {Xia}}, \bibinfo {author} {\bibfnamefont {Y.}~\bibnamefont {Xu}}, \bibinfo
  {author} {\bibfnamefont {P.}~\bibnamefont {Malinowski}}, \bibinfo {author}
  {\bibfnamefont {X.~D.}\ \bibnamefont {Xu}}, \bibinfo {author} {\bibfnamefont
  {J.~H.}\ \bibnamefont {Chu}}, \bibinfo {author} {\bibfnamefont
  {G.}~\bibnamefont {Li}}, \bibinfo {author} {\bibfnamefont {Y.~F.}\
  \bibnamefont {Guo}}, \bibinfo {author} {\bibfnamefont {Z.~K.}\ \bibnamefont
  {Liu}}, \bibinfo {author} {\bibfnamefont {L.~X.}\ \bibnamefont {Yang}},\ and\
  \bibinfo {author} {\bibfnamefont {Y.~L.}\ \bibnamefont {Chen}},\ }\bibfield
  {title} {\bibinfo {title} {{Signature for non-Stoner ferromagnetism in the
  van der Waals ferromagnet Fe$_3$GeTe$_2$}},\ }\href
  {https://doi.org/10.1103/PHYSREVB.101.201104/FIGURES/4/MEDIUM} {\bibfield
  {journal} {\bibinfo  {journal} {Phys. Rev. B}\ }\textbf {\bibinfo {volume}
  {101}},\ \bibinfo {pages} {201104} (\bibinfo {year} {2020})}\BibitemShut
  {NoStop}%
\bibitem [{\citenamefont {Bettler}\ \emph {et~al.}(2019)\citenamefont
  {Bettler}, \citenamefont {Landolt}, \citenamefont {Aksoy}, \citenamefont
  {Yan}, \citenamefont {Gvasaliya}, \citenamefont {Qiu}, \citenamefont
  {Ressouche}, \citenamefont {Beauvois}, \citenamefont {Raymond}, \citenamefont
  {Ponomaryov}, \citenamefont {Zvyagin},\ and\ \citenamefont
  {Zheludev}}]{Bettler2019MagneticPbVO}%
  \BibitemOpen
  \bibfield  {author} {\bibinfo {author} {\bibfnamefont {S.}~\bibnamefont
  {Bettler}}, \bibinfo {author} {\bibfnamefont {F.}~\bibnamefont {Landolt}},
  \bibinfo {author} {\bibfnamefont {O.~M.}\ \bibnamefont {Aksoy}}, \bibinfo
  {author} {\bibfnamefont {Z.}~\bibnamefont {Yan}}, \bibinfo {author}
  {\bibfnamefont {S.}~\bibnamefont {Gvasaliya}}, \bibinfo {author}
  {\bibfnamefont {Y.}~\bibnamefont {Qiu}}, \bibinfo {author} {\bibfnamefont
  {E.}~\bibnamefont {Ressouche}}, \bibinfo {author} {\bibfnamefont
  {K.}~\bibnamefont {Beauvois}}, \bibinfo {author} {\bibfnamefont
  {S.}~\bibnamefont {Raymond}}, \bibinfo {author} {\bibfnamefont {A.~N.}\
  \bibnamefont {Ponomaryov}}, \bibinfo {author} {\bibfnamefont {S.~A.}\
  \bibnamefont {Zvyagin}},\ and\ \bibinfo {author} {\bibfnamefont
  {A.}~\bibnamefont {Zheludev}},\ }\bibfield  {title} {\bibinfo {title}
  {Magnetic structure and spin waves in the frustrated ferro-antiferromagnet
  {Pb}$_{2}\mathrm{VO}{({\mathrm{PO}}_{4})}_{2}$},\ }\href
  {https://doi.org/10.1103/PhysRevB.99.184437} {\bibfield  {journal} {\bibinfo
  {journal} {Phys. Rev. B}\ }\textbf {\bibinfo {volume} {99}},\ \bibinfo
  {pages} {184437} (\bibinfo {year} {2019})}\BibitemShut {NoStop}%
\bibitem [{\citenamefont {Bhartiya}\ \emph {et~al.}(2021)\citenamefont
  {Bhartiya}, \citenamefont {Hayashida}, \citenamefont {Povarov}, \citenamefont
  {Yan}, \citenamefont {Qiu}, \citenamefont {Raymond},\ and\ \citenamefont
  {Zheludev}}]{Bhartiya2021InelasticBaVO}%
  \BibitemOpen
  \bibfield  {author} {\bibinfo {author} {\bibfnamefont {V.~K.}\ \bibnamefont
  {Bhartiya}}, \bibinfo {author} {\bibfnamefont {S.}~\bibnamefont {Hayashida}},
  \bibinfo {author} {\bibfnamefont {K.~Y.}\ \bibnamefont {Povarov}}, \bibinfo
  {author} {\bibfnamefont {Z.}~\bibnamefont {Yan}}, \bibinfo {author}
  {\bibfnamefont {Y.}~\bibnamefont {Qiu}}, \bibinfo {author} {\bibfnamefont
  {S.}~\bibnamefont {Raymond}},\ and\ \bibinfo {author} {\bibfnamefont
  {A.}~\bibnamefont {Zheludev}},\ }\bibfield  {title} {\bibinfo {title}
  {Inelastic neutron scattering determination of the spin hamiltonian for
  $\mathrm{BaCdVO}{({\mathrm{PO}}_{4})}_{2}$},\ }\href
  {https://doi.org/10.1103/PhysRevB.103.144402} {\bibfield  {journal} {\bibinfo
   {journal} {Phys. Rev. B}\ }\textbf {\bibinfo {volume} {103}},\ \bibinfo
  {pages} {144402} (\bibinfo {year} {2021})}\BibitemShut {NoStop}%
\bibitem [{\citenamefont {Pelliciari}\ \emph {et~al.}(2021)\citenamefont
  {Pelliciari}, \citenamefont {Lee}, \citenamefont {Gilmore}, \citenamefont
  {Li}, \citenamefont {Gu}, \citenamefont {Barbour}, \citenamefont {Jarrige},
  \citenamefont {Ahn}, \citenamefont {Walker},\ and\ \citenamefont
  {Bisogni}}]{Pelliciari2021TuningConfinement}%
  \BibitemOpen
  \bibfield  {author} {\bibinfo {author} {\bibfnamefont {J.}~\bibnamefont
  {Pelliciari}}, \bibinfo {author} {\bibfnamefont {S.}~\bibnamefont {Lee}},
  \bibinfo {author} {\bibfnamefont {K.}~\bibnamefont {Gilmore}}, \bibinfo
  {author} {\bibfnamefont {J.}~\bibnamefont {Li}}, \bibinfo {author}
  {\bibfnamefont {Y.}~\bibnamefont {Gu}}, \bibinfo {author} {\bibfnamefont
  {A.}~\bibnamefont {Barbour}}, \bibinfo {author} {\bibfnamefont
  {I.}~\bibnamefont {Jarrige}}, \bibinfo {author} {\bibfnamefont {C.~H.}\
  \bibnamefont {Ahn}}, \bibinfo {author} {\bibfnamefont {F.~J.}\ \bibnamefont
  {Walker}},\ and\ \bibinfo {author} {\bibfnamefont {V.}~\bibnamefont
  {Bisogni}},\ }\bibfield  {title} {\bibinfo {title} {Tuning spin excitations
  in magnetic films by confinement},\ }\href
  {https://doi.org/10.1038/s41563-020-00878-0} {\bibfield  {journal} {\bibinfo
  {journal} {Nature Materials}\ }\textbf {\bibinfo {volume} {20}},\ \bibinfo
  {pages} {188} (\bibinfo {year} {2021})}\BibitemShut {NoStop}%
\bibitem [{\citenamefont {Abbamonte}\ \emph {et~al.}(2004)\citenamefont
  {Abbamonte}, \citenamefont {Blumberg}, \citenamefont {Rusydi}, \citenamefont
  {Gozar}, \citenamefont {Evans}, \citenamefont {Siegrist}, \citenamefont
  {Venema}, \citenamefont {Eisaki}, \citenamefont {Isaacs},\ and\ \citenamefont
  {Sawatzky}}]{Abbamonte2004CrystallizationSr14Cu24O41}%
  \BibitemOpen
  \bibfield  {author} {\bibinfo {author} {\bibfnamefont {P.}~\bibnamefont
  {Abbamonte}}, \bibinfo {author} {\bibfnamefont {G.}~\bibnamefont {Blumberg}},
  \bibinfo {author} {\bibfnamefont {A.}~\bibnamefont {Rusydi}}, \bibinfo
  {author} {\bibfnamefont {A.}~\bibnamefont {Gozar}}, \bibinfo {author}
  {\bibfnamefont {P.~G.}\ \bibnamefont {Evans}}, \bibinfo {author}
  {\bibfnamefont {T.}~\bibnamefont {Siegrist}}, \bibinfo {author}
  {\bibfnamefont {L.}~\bibnamefont {Venema}}, \bibinfo {author} {\bibfnamefont
  {H.}~\bibnamefont {Eisaki}}, \bibinfo {author} {\bibfnamefont {E.~D.}\
  \bibnamefont {Isaacs}},\ and\ \bibinfo {author} {\bibfnamefont {G.~A.}\
  \bibnamefont {Sawatzky}},\ }\bibfield  {title} {\bibinfo {title}
  {{Crystallization of charge holes in the spin ladder of
  Sr$_{14}$Cu$_{24}$O$_{41}$}},\ }\href {https://doi.org/10.1038/nature02925}
  {\bibfield  {journal} {\bibinfo  {journal} {Nature}\ }\textbf {\bibinfo
  {volume} {431}},\ \bibinfo {pages} {1078} (\bibinfo {year}
  {2004})}\BibitemShut {NoStop}%
\bibitem [{\citenamefont {Chen}\ \emph {et~al.}(2016)\citenamefont {Chen},
  \citenamefont {Thampy}, \citenamefont {Mazzoli}, \citenamefont {Barbour},
  \citenamefont {Miao}, \citenamefont {Gu}, \citenamefont {Cao}, \citenamefont
  {Tranquada}, \citenamefont {Dean},\ and\ \citenamefont
  {Wilkins}}]{Chen2016Remarkable4}%
  \BibitemOpen
  \bibfield  {author} {\bibinfo {author} {\bibfnamefont {X.~M.}\ \bibnamefont
  {Chen}}, \bibinfo {author} {\bibfnamefont {V.}~\bibnamefont {Thampy}},
  \bibinfo {author} {\bibfnamefont {C.}~\bibnamefont {Mazzoli}}, \bibinfo
  {author} {\bibfnamefont {A.~M.}\ \bibnamefont {Barbour}}, \bibinfo {author}
  {\bibfnamefont {H.}~\bibnamefont {Miao}}, \bibinfo {author} {\bibfnamefont
  {G.~D.}\ \bibnamefont {Gu}}, \bibinfo {author} {\bibfnamefont
  {Y.}~\bibnamefont {Cao}}, \bibinfo {author} {\bibfnamefont {J.~M.}\
  \bibnamefont {Tranquada}}, \bibinfo {author} {\bibfnamefont {M.~P.~M.}\
  \bibnamefont {Dean}},\ and\ \bibinfo {author} {\bibfnamefont {S.~B.}\
  \bibnamefont {Wilkins}},\ }\bibfield  {title} {\bibinfo {title} {Remarkable
  stability of charge density wave order in
  {La}$_{1.875}${Ba}$_{0.125}${CuO}$_{4}$},\ }\href
  {https://doi.org/10.1103/PhysRevLett.117.167001} {\bibfield  {journal}
  {\bibinfo  {journal} {Phys. Rev. Lett.}\ }\textbf {\bibinfo {volume} {117}},\
  \bibinfo {pages} {167001} (\bibinfo {year} {2016})}\BibitemShut {NoStop}%
\bibitem [{\citenamefont {Shen}\ \emph {et~al.}(2021)\citenamefont {Shen},
  \citenamefont {Fabbris}, \citenamefont {Miao}, \citenamefont {Cao},
  \citenamefont {Meyers}, \citenamefont {Mazzone}, \citenamefont {Assefa},
  \citenamefont {Chen}, \citenamefont {Kisslinger}, \citenamefont
  {Prabhakaran}, \citenamefont {Boothroyd}, \citenamefont {Tranquada},
  \citenamefont {Hu}, \citenamefont {Barbour}, \citenamefont {Wilkins},
  \citenamefont {Mazzoli}, \citenamefont {Robinson},\ and\ \citenamefont
  {Dean}}]{Shen2021ChargeNickelates}%
  \BibitemOpen
  \bibfield  {author} {\bibinfo {author} {\bibfnamefont {Y.}~\bibnamefont
  {Shen}}, \bibinfo {author} {\bibfnamefont {G.}~\bibnamefont {Fabbris}},
  \bibinfo {author} {\bibfnamefont {H.}~\bibnamefont {Miao}}, \bibinfo {author}
  {\bibfnamefont {Y.}~\bibnamefont {Cao}}, \bibinfo {author} {\bibfnamefont
  {D.}~\bibnamefont {Meyers}}, \bibinfo {author} {\bibfnamefont {D.~G.}\
  \bibnamefont {Mazzone}}, \bibinfo {author} {\bibfnamefont {T.~A.}\
  \bibnamefont {Assefa}}, \bibinfo {author} {\bibfnamefont {X.~M.}\
  \bibnamefont {Chen}}, \bibinfo {author} {\bibfnamefont {K.}~\bibnamefont
  {Kisslinger}}, \bibinfo {author} {\bibfnamefont {D.}~\bibnamefont
  {Prabhakaran}}, \bibinfo {author} {\bibfnamefont {A.~T.}\ \bibnamefont
  {Boothroyd}}, \bibinfo {author} {\bibfnamefont {J.~M.}\ \bibnamefont
  {Tranquada}}, \bibinfo {author} {\bibfnamefont {W.}~\bibnamefont {Hu}},
  \bibinfo {author} {\bibfnamefont {A.~M.}\ \bibnamefont {Barbour}}, \bibinfo
  {author} {\bibfnamefont {S.~B.}\ \bibnamefont {Wilkins}}, \bibinfo {author}
  {\bibfnamefont {C.}~\bibnamefont {Mazzoli}}, \bibinfo {author} {\bibfnamefont
  {I.~K.}\ \bibnamefont {Robinson}},\ and\ \bibinfo {author} {\bibfnamefont
  {M.~P.~M.}\ \bibnamefont {Dean}},\ }\bibfield  {title} {\bibinfo {title}
  {Charge condensation and lattice coupling drives stripe formation in
  nickelates},\ }\href {https://doi.org/10.1103/PhysRevLett.126.177601}
  {\bibfield  {journal} {\bibinfo  {journal} {Phys. Rev. Lett.}\ }\textbf
  {\bibinfo {volume} {126}},\ \bibinfo {pages} {177601} (\bibinfo {year}
  {2021})}\BibitemShut {NoStop}%
\bibitem [{\citenamefont {May}\ \emph {et~al.}(2016)\citenamefont {May},
  \citenamefont {Calder}, \citenamefont {Cantoni}, \citenamefont {Cao},\ and\
  \citenamefont {Mcguire}}]{MayPRB2016_Fe3GT}%
  \BibitemOpen
  \bibfield  {author} {\bibinfo {author} {\bibfnamefont {A.~F.}\ \bibnamefont
  {May}}, \bibinfo {author} {\bibfnamefont {S.}~\bibnamefont {Calder}},
  \bibinfo {author} {\bibfnamefont {C.}~\bibnamefont {Cantoni}}, \bibinfo
  {author} {\bibfnamefont {H.}~\bibnamefont {Cao}},\ and\ \bibinfo {author}
  {\bibfnamefont {M.~A.}\ \bibnamefont {Mcguire}},\ }\bibfield  {title}
  {\bibinfo {title} {{Magnetic structure and phase stability of the van der
  Waals bonded ferromagnet {Fe}$_{3-x}${GeTe}$_{2}$}},\ }\href
  {https://doi.org/10.1103/PhysRevB.93.014411} {\bibfield  {journal} {\bibinfo
  {journal} {Phys. Rev. B}\ }\textbf {\bibinfo {volume} {93}},\ \bibinfo
  {pages} {14411} (\bibinfo {year} {2016})}\BibitemShut {NoStop}%
\bibitem [{\citenamefont {Yamagami}\ \emph {et~al.}(2022)\citenamefont
  {Yamagami}, \citenamefont {Fujisawa}, \citenamefont {Pardo-Almanza},
  \citenamefont {Smith}, \citenamefont {Sumida}, \citenamefont {Takeda},\ and\
  \citenamefont {Okada}}]{Yamagami2022EnhancedFe5GT}%
  \BibitemOpen
  \bibfield  {author} {\bibinfo {author} {\bibfnamefont {K.}~\bibnamefont
  {Yamagami}}, \bibinfo {author} {\bibfnamefont {Y.}~\bibnamefont {Fujisawa}},
  \bibinfo {author} {\bibfnamefont {M.}~\bibnamefont {Pardo-Almanza}}, \bibinfo
  {author} {\bibfnamefont {B.~R.~M.}\ \bibnamefont {Smith}}, \bibinfo {author}
  {\bibfnamefont {K.}~\bibnamefont {Sumida}}, \bibinfo {author} {\bibfnamefont
  {Y.}~\bibnamefont {Takeda}},\ and\ \bibinfo {author} {\bibfnamefont
  {Y.}~\bibnamefont {Okada}},\ }\bibfield  {title} {\bibinfo {title} {{Enhanced
  d-p hybridization intertwined with anomalous ground state formation in the
  van der Waals itinerant magnet {Fe}$_5${GeTe}$_2$}},\ }\href
  {https://doi.org/10.1103/PhysRevB.106.045137} {\bibfield  {journal} {\bibinfo
   {journal} {Phys. Rev. B}\ }\textbf {\bibinfo {volume} {106}},\ \bibinfo
  {pages} {45137} (\bibinfo {year} {2022})}\BibitemShut {NoStop}%
\bibitem [{\citenamefont {Calder}\ \emph {et~al.}(2019)\citenamefont {Calder},
  \citenamefont {Kolesnikov},\ and\ \citenamefont {May}}]{Calder2019}%
  \BibitemOpen
  \bibfield  {author} {\bibinfo {author} {\bibfnamefont {S.}~\bibnamefont
  {Calder}}, \bibinfo {author} {\bibfnamefont {A.~I.}\ \bibnamefont
  {Kolesnikov}},\ and\ \bibinfo {author} {\bibfnamefont {A.~F.}\ \bibnamefont
  {May}},\ }\bibfield  {title} {\bibinfo {title} {Magnetic excitations in the
  quasi-two-dimensional ferromagnet
  $\mathrm{Fe}_{3\ensuremath{-}x}\mathrm{GeTe}_2$ measured with inelastic
  neutron scattering},\ }\href {https://doi.org/10.1103/PhysRevB.99.094423}
  {\bibfield  {journal} {\bibinfo  {journal} {Phys. Rev. B}\ }\textbf {\bibinfo
  {volume} {99}},\ \bibinfo {pages} {094423} (\bibinfo {year}
  {2019})}\BibitemShut {NoStop}%
\bibitem [{\citenamefont {Li}\ \emph {et~al.}(2023)\citenamefont {Li},
  \citenamefont {Gu}, \citenamefont {Takahashi}, \citenamefont {Higashi},
  \citenamefont {Kim}, \citenamefont {Cheng}, \citenamefont {Yang},
  \citenamefont {Kune\ifmmode~\check{s}\else \v{s}\fi{}}, \citenamefont
  {Pelliciari}, \citenamefont {Hariki},\ and\ \citenamefont
  {Bisogni}}]{Li_PRX_2023}%
  \BibitemOpen
  \bibfield  {author} {\bibinfo {author} {\bibfnamefont {J.}~\bibnamefont
  {Li}}, \bibinfo {author} {\bibfnamefont {Y.}~\bibnamefont {Gu}}, \bibinfo
  {author} {\bibfnamefont {Y.}~\bibnamefont {Takahashi}}, \bibinfo {author}
  {\bibfnamefont {K.}~\bibnamefont {Higashi}}, \bibinfo {author} {\bibfnamefont
  {T.}~\bibnamefont {Kim}}, \bibinfo {author} {\bibfnamefont {Y.}~\bibnamefont
  {Cheng}}, \bibinfo {author} {\bibfnamefont {F.}~\bibnamefont {Yang}},
  \bibinfo {author} {\bibfnamefont {J.}~\bibnamefont
  {Kune\ifmmode~\check{s}\else \v{s}\fi{}}}, \bibinfo {author} {\bibfnamefont
  {J.}~\bibnamefont {Pelliciari}}, \bibinfo {author} {\bibfnamefont
  {A.}~\bibnamefont {Hariki}},\ and\ \bibinfo {author} {\bibfnamefont
  {V.}~\bibnamefont {Bisogni}},\ }\bibfield  {title} {\bibinfo {title} {Single-
  and multimagnon dynamics in antiferromagnetic $\alpha -${Fe}$_2${O}$_3$ thin
  films},\ }\href {https://doi.org/10.1103/PhysRevX.13.011012} {\bibfield
  {journal} {\bibinfo  {journal} {Phys. Rev. X}\ }\textbf {\bibinfo {volume}
  {13}},\ \bibinfo {pages} {011012} (\bibinfo {year} {2023})}\BibitemShut
  {NoStop}%
\bibitem [{\citenamefont {Elnaggar}\ \emph {et~al.}(2023)\citenamefont
  {Elnaggar}, \citenamefont {Nag}, \citenamefont {Haverkort}, \citenamefont
  {Garcia-Fernandez}, \citenamefont {Walters}, \citenamefont {Wang},
  \citenamefont {Zhou},\ and\ \citenamefont
  {de~Groot}}]{Elnaggar_NatComm_2023}%
  \BibitemOpen
  \bibfield  {author} {\bibinfo {author} {\bibfnamefont {H.}~\bibnamefont
  {Elnaggar}}, \bibinfo {author} {\bibfnamefont {A.}~\bibnamefont {Nag}},
  \bibinfo {author} {\bibfnamefont {M.~W.}\ \bibnamefont {Haverkort}}, \bibinfo
  {author} {\bibfnamefont {M.}~\bibnamefont {Garcia-Fernandez}}, \bibinfo
  {author} {\bibfnamefont {A.}~\bibnamefont {Walters}}, \bibinfo {author}
  {\bibfnamefont {R.-P.}\ \bibnamefont {Wang}}, \bibinfo {author}
  {\bibfnamefont {K.-J.}\ \bibnamefont {Zhou}},\ and\ \bibinfo {author}
  {\bibfnamefont {F.}~\bibnamefont {de~Groot}},\ }\bibfield  {title} {\bibinfo
  {title} {Magnetic excitations beyond the single- and double-magnons},\ }\href
  {https://doi.org/10.1038/s41467-023-38341-8} {\bibfield  {journal} {\bibinfo
  {journal} {Nature Communications}\ }\textbf {\bibinfo {volume} {14}},\
  \bibinfo {pages} {2749} (\bibinfo {year} {2023})}\BibitemShut {NoStop}%
\bibitem [{\citenamefont {Xie}\ \emph {et~al.}(2018)\citenamefont {Xie},
  \citenamefont {Wei}, \citenamefont {Gong}, \citenamefont {Fennell},
  \citenamefont {Stuhr}, \citenamefont {Kajimoto}, \citenamefont {Ikeuchi},
  \citenamefont {Li}, \citenamefont {Hu},\ and\ \citenamefont
  {Luo}}]{Xie_PRL_2018}%
  \BibitemOpen
  \bibfield  {author} {\bibinfo {author} {\bibfnamefont {T.}~\bibnamefont
  {Xie}}, \bibinfo {author} {\bibfnamefont {Y.}~\bibnamefont {Wei}}, \bibinfo
  {author} {\bibfnamefont {D.}~\bibnamefont {Gong}}, \bibinfo {author}
  {\bibfnamefont {T.}~\bibnamefont {Fennell}}, \bibinfo {author} {\bibfnamefont
  {U.}~\bibnamefont {Stuhr}}, \bibinfo {author} {\bibfnamefont
  {R.}~\bibnamefont {Kajimoto}}, \bibinfo {author} {\bibfnamefont
  {K.}~\bibnamefont {Ikeuchi}}, \bibinfo {author} {\bibfnamefont
  {S.}~\bibnamefont {Li}}, \bibinfo {author} {\bibfnamefont {J.}~\bibnamefont
  {Hu}},\ and\ \bibinfo {author} {\bibfnamefont {H.}~\bibnamefont {Luo}},\
  }\bibfield  {title} {\bibinfo {title} {Odd and even modes of neutron spin
  resonance in the bilayer iron-based superconductor {CaKFe}$_{4}${As}$_{4}$},\
  }\href {https://doi.org/10.1103/PhysRevLett.120.267003} {\bibfield  {journal}
  {\bibinfo  {journal} {Phys. Rev. Lett.}\ }\textbf {\bibinfo {volume} {120}},\
  \bibinfo {pages} {267003} (\bibinfo {year} {2018})}\BibitemShut {NoStop}%
\bibitem [{\citenamefont {Chen}\ \emph {et~al.}(2020)\citenamefont {Chen},
  \citenamefont {Krivenko}, \citenamefont {Stone}, \citenamefont {Kolesnikov},
  \citenamefont {Wolf}, \citenamefont {Reznik}, \citenamefont {Bedell},
  \citenamefont {Lechermann},\ and\ \citenamefont
  {Wilson}}]{Chen_NatComm_2020}%
  \BibitemOpen
  \bibfield  {author} {\bibinfo {author} {\bibfnamefont {X.}~\bibnamefont
  {Chen}}, \bibinfo {author} {\bibfnamefont {I.}~\bibnamefont {Krivenko}},
  \bibinfo {author} {\bibfnamefont {M.~B.}\ \bibnamefont {Stone}}, \bibinfo
  {author} {\bibfnamefont {A.~I.}\ \bibnamefont {Kolesnikov}}, \bibinfo
  {author} {\bibfnamefont {T.}~\bibnamefont {Wolf}}, \bibinfo {author}
  {\bibfnamefont {D.}~\bibnamefont {Reznik}}, \bibinfo {author} {\bibfnamefont
  {K.~S.}\ \bibnamefont {Bedell}}, \bibinfo {author} {\bibfnamefont
  {F.}~\bibnamefont {Lechermann}},\ and\ \bibinfo {author} {\bibfnamefont
  {S.~D.}\ \bibnamefont {Wilson}},\ }\bibfield  {title} {\bibinfo {title}
  {Unconventional hund metal in a weak itinerant ferromagnet},\ }\href
  {https://www.nature.com/articles/s41467-020-16868-4} {\bibfield  {journal}
  {\bibinfo  {journal} {Nature Communications}\ }\textbf {\bibinfo {volume}
  {11}},\ \bibinfo {pages} {3076} (\bibinfo {year} {2020})}\BibitemShut
  {NoStop}%
\bibitem [{\citenamefont {Song}\ \emph {et~al.}(2021)\citenamefont {Song},
  \citenamefont {Wang}, \citenamefont {Cao}, \citenamefont {Yamani},
  \citenamefont {Xu}, \citenamefont {Sheng}, \citenamefont {L{\"o}ser},
  \citenamefont {Qiu}, \citenamefont {Yang}, \citenamefont {Birgeneau},\ and\
  \citenamefont {Dai}}]{Song_npjQM_2021}%
  \BibitemOpen
  \bibfield  {author} {\bibinfo {author} {\bibfnamefont {Y.}~\bibnamefont
  {Song}}, \bibinfo {author} {\bibfnamefont {W.}~\bibnamefont {Wang}}, \bibinfo
  {author} {\bibfnamefont {C.}~\bibnamefont {Cao}}, \bibinfo {author}
  {\bibfnamefont {Z.}~\bibnamefont {Yamani}}, \bibinfo {author} {\bibfnamefont
  {Y.}~\bibnamefont {Xu}}, \bibinfo {author} {\bibfnamefont {Y.}~\bibnamefont
  {Sheng}}, \bibinfo {author} {\bibfnamefont {W.}~\bibnamefont {L{\"o}ser}},
  \bibinfo {author} {\bibfnamefont {Y.}~\bibnamefont {Qiu}}, \bibinfo {author}
  {\bibfnamefont {Y.-f.}\ \bibnamefont {Yang}}, \bibinfo {author}
  {\bibfnamefont {R.~J.}\ \bibnamefont {Birgeneau}},\ and\ \bibinfo {author}
  {\bibfnamefont {P.}~\bibnamefont {Dai}},\ }\bibfield  {title} {\bibinfo
  {title} {High-energy magnetic excitations from heavy quasiparticles in
  {CeCu}$_2${Si}$_2$},\ }\href
  {https://www.nature.com/articles/s41535-021-00358-x} {\bibfield  {journal}
  {\bibinfo  {journal} {npj Quantum Materials}\ }\textbf {\bibinfo {volume}
  {6}},\ \bibinfo {pages} {60} (\bibinfo {year} {2021})}\BibitemShut {NoStop}%
\bibitem [{\citenamefont {Anisimov}\ \emph {et~al.}(2002)\citenamefont
  {Anisimov}, \citenamefont {Nekrasov}, \citenamefont {Kondakov}, \citenamefont
  {Rice},\ and\ \citenamefont {Sigrist}}]{Anisimov2002}%
  \BibitemOpen
  \bibfield  {author} {\bibinfo {author} {\bibfnamefont {V.~I.}\ \bibnamefont
  {Anisimov}}, \bibinfo {author} {\bibfnamefont {I.~A.}\ \bibnamefont
  {Nekrasov}}, \bibinfo {author} {\bibfnamefont {D.~E.}\ \bibnamefont
  {Kondakov}}, \bibinfo {author} {\bibfnamefont {T.~M.}\ \bibnamefont {Rice}},\
  and\ \bibinfo {author} {\bibfnamefont {M.}~\bibnamefont {Sigrist}},\
  }\bibfield  {title} {\bibinfo {title} {Orbital-selective mott-insulator
  transition in {Ca}$_{2-x}${Sr}$_x${RuO}$_{4}$},\ }\href
  {https://doi.org/10.1140/epjb/e20020021} {\bibfield  {journal} {\bibinfo
  {journal} {The European Physical Journal B - Condensed Matter and Complex
  Systems}\ }\textbf {\bibinfo {volume} {25}},\ \bibinfo {pages} {191}
  (\bibinfo {year} {2002})}\BibitemShut {NoStop}%
\bibitem [{\citenamefont {Kim}\ \emph {et~al.}(2022)\citenamefont {Kim},
  \citenamefont {Kwon}, \citenamefont {Kim}, \citenamefont {Kim}, \citenamefont
  {Chung}, \citenamefont {Ryu}, \citenamefont {Jung}, \citenamefont {Kim},
  \citenamefont {Song}, \citenamefont {Denlinger}, \citenamefont {Han},
  \citenamefont {Yoshida}, \citenamefont {Mizokawa}, \citenamefont {Kyung},\
  and\ \citenamefont {Kim}}]{Kim_QM_2022}%
  \BibitemOpen
  \bibfield  {author} {\bibinfo {author} {\bibfnamefont {M.}~\bibnamefont
  {Kim}}, \bibinfo {author} {\bibfnamefont {J.}~\bibnamefont {Kwon}}, \bibinfo
  {author} {\bibfnamefont {C.~H.}\ \bibnamefont {Kim}}, \bibinfo {author}
  {\bibfnamefont {Y.}~\bibnamefont {Kim}}, \bibinfo {author} {\bibfnamefont
  {D.}~\bibnamefont {Chung}}, \bibinfo {author} {\bibfnamefont
  {H.}~\bibnamefont {Ryu}}, \bibinfo {author} {\bibfnamefont {J.}~\bibnamefont
  {Jung}}, \bibinfo {author} {\bibfnamefont {B.~S.}\ \bibnamefont {Kim}},
  \bibinfo {author} {\bibfnamefont {D.}~\bibnamefont {Song}}, \bibinfo {author}
  {\bibfnamefont {J.~D.}\ \bibnamefont {Denlinger}}, \bibinfo {author}
  {\bibfnamefont {M.}~\bibnamefont {Han}}, \bibinfo {author} {\bibfnamefont
  {Y.}~\bibnamefont {Yoshida}}, \bibinfo {author} {\bibfnamefont
  {T.}~\bibnamefont {Mizokawa}}, \bibinfo {author} {\bibfnamefont
  {W.}~\bibnamefont {Kyung}},\ and\ \bibinfo {author} {\bibfnamefont
  {C.}~\bibnamefont {Kim}},\ }\bibfield  {title} {\bibinfo {title} {Signature
  of kondo hybridisation with an orbital-selective mott phase in 4d
  {Ca}$_{2-x}${Sr}$_x${RuO}$_{4}$},\ }\href
  {https://www.nature.com/articles/s41535-022-00471-5} {\bibfield  {journal}
  {\bibinfo  {journal} {npj Quantum Materials}\ }\textbf {\bibinfo {volume}
  {7}},\ \bibinfo {pages} {59} (\bibinfo {year} {2022})}\BibitemShut {NoStop}%
\bibitem [{\citenamefont {de' Medici}\ \emph {et~al.}(2005)\citenamefont {de'
  Medici}, \citenamefont {Georges}, \citenamefont {Kotliar},\ and\
  \citenamefont {Biermann}}]{Medici_PRL_2005}%
  \BibitemOpen
  \bibfield  {author} {\bibinfo {author} {\bibfnamefont {L.}~\bibnamefont {de'
  Medici}}, \bibinfo {author} {\bibfnamefont {A.}~\bibnamefont {Georges}},
  \bibinfo {author} {\bibfnamefont {G.}~\bibnamefont {Kotliar}},\ and\ \bibinfo
  {author} {\bibfnamefont {S.}~\bibnamefont {Biermann}},\ }\bibfield  {title}
  {\bibinfo {title} {Mott transition and kondo screening in $f$-electron
  metals},\ }\href {https://doi.org/10.1103/PhysRevLett.95.066402} {\bibfield
  {journal} {\bibinfo  {journal} {Phys. Rev. Lett.}\ }\textbf {\bibinfo
  {volume} {95}},\ \bibinfo {pages} {066402} (\bibinfo {year}
  {2005})}\BibitemShut {NoStop}%
\bibitem [{\citenamefont {de' Medici}\ \emph {et~al.}(2009)\citenamefont {de'
  Medici}, \citenamefont {Hassan}, \citenamefont {Capone},\ and\ \citenamefont
  {Dai}}]{Medici_PRL_2009}%
  \BibitemOpen
  \bibfield  {author} {\bibinfo {author} {\bibfnamefont {L.}~\bibnamefont {de'
  Medici}}, \bibinfo {author} {\bibfnamefont {S.~R.}\ \bibnamefont {Hassan}},
  \bibinfo {author} {\bibfnamefont {M.}~\bibnamefont {Capone}},\ and\ \bibinfo
  {author} {\bibfnamefont {X.}~\bibnamefont {Dai}},\ }\bibfield  {title}
  {\bibinfo {title} {Orbital-selective mott transition out of band degeneracy
  lifting},\ }\href {https://doi.org/10.1103/PhysRevLett.102.126401} {\bibfield
   {journal} {\bibinfo  {journal} {Phys. Rev. Lett.}\ }\textbf {\bibinfo
  {volume} {102}},\ \bibinfo {pages} {126401} (\bibinfo {year}
  {2009})}\BibitemShut {NoStop}%
\bibitem [{\citenamefont {Yamagami}\ \emph {et~al.}(2021)\citenamefont
  {Yamagami}, \citenamefont {Fujisawa}, \citenamefont {Driesen}, \citenamefont
  {Hsu}, \citenamefont {Kawaguchi}, \citenamefont {Tanaka}, \citenamefont
  {Kondo}, \citenamefont {Zhang}, \citenamefont {Wadati}, \citenamefont
  {Araki}, \citenamefont {Takeda}, \citenamefont {Takeda}, \citenamefont
  {Muro}, \citenamefont {Chuang}, \citenamefont {Niimi}, \citenamefont
  {Kuroda}, \citenamefont {Kobayashi},\ and\ \citenamefont
  {Okada}}]{Yamagami2021giant}%
  \BibitemOpen
  \bibfield  {author} {\bibinfo {author} {\bibfnamefont {K.}~\bibnamefont
  {Yamagami}}, \bibinfo {author} {\bibfnamefont {Y.}~\bibnamefont {Fujisawa}},
  \bibinfo {author} {\bibfnamefont {B.}~\bibnamefont {Driesen}}, \bibinfo
  {author} {\bibfnamefont {C.~H.}\ \bibnamefont {Hsu}}, \bibinfo {author}
  {\bibfnamefont {K.}~\bibnamefont {Kawaguchi}}, \bibinfo {author}
  {\bibfnamefont {H.}~\bibnamefont {Tanaka}}, \bibinfo {author} {\bibfnamefont
  {T.}~\bibnamefont {Kondo}}, \bibinfo {author} {\bibfnamefont
  {Y.}~\bibnamefont {Zhang}}, \bibinfo {author} {\bibfnamefont
  {H.}~\bibnamefont {Wadati}}, \bibinfo {author} {\bibfnamefont
  {K.}~\bibnamefont {Araki}}, \bibinfo {author} {\bibfnamefont
  {T.}~\bibnamefont {Takeda}}, \bibinfo {author} {\bibfnamefont
  {Y.}~\bibnamefont {Takeda}}, \bibinfo {author} {\bibfnamefont
  {T.}~\bibnamefont {Muro}}, \bibinfo {author} {\bibfnamefont {F.~C.}\
  \bibnamefont {Chuang}}, \bibinfo {author} {\bibfnamefont {Y.}~\bibnamefont
  {Niimi}}, \bibinfo {author} {\bibfnamefont {K.}~\bibnamefont {Kuroda}},
  \bibinfo {author} {\bibfnamefont {M.}~\bibnamefont {Kobayashi}},\ and\
  \bibinfo {author} {\bibfnamefont {Y.}~\bibnamefont {Okada}},\ }\bibfield
  {title} {\bibinfo {title} {Itinerant ferromagnetism mediated by giant spin
  polarization of the metallic ligand band in the van der waals magnet
  {Fe}$_{5}${Ge}{Te}$_{2}$},\ }\href
  {https://doi.org/10.1103/PhysRevB.103.L060403} {\bibfield  {journal}
  {\bibinfo  {journal} {Phys. Rev. B}\ }\textbf {\bibinfo {volume} {103}},\
  \bibinfo {pages} {L060403} (\bibinfo {year} {2021})}\BibitemShut {NoStop}%
\bibitem [{\citenamefont {Wang}\ and\ \citenamefont
  {Zhang}(2023)}]{Wang_PRB_2023}%
  \BibitemOpen
  \bibfield  {author} {\bibinfo {author} {\bibfnamefont {F.}~\bibnamefont
  {Wang}}\ and\ \bibinfo {author} {\bibfnamefont {H.}~\bibnamefont {Zhang}},\
  }\bibfield  {title} {\bibinfo {title} {Flat bands and magnetism in
  {Fe}$_{4}${GeTe}$_{2}$ and {Fe}$_{5}${GeTe}$_{2}$ due to bipartite crystal
  lattices},\ }\href {https://doi.org/10.1103/PhysRevB.108.195140} {\bibfield
  {journal} {\bibinfo  {journal} {Phys. Rev. B}\ }\textbf {\bibinfo {volume}
  {108}},\ \bibinfo {pages} {195140} (\bibinfo {year} {2023})}\BibitemShut
  {NoStop}%
\bibitem [{\citenamefont {Revelli}\ \emph {et~al.}(2020)\citenamefont
  {Revelli}, \citenamefont {Moretti~Sala}, \citenamefont {Monaco},
  \citenamefont {Hickey}, \citenamefont {Becker}, \citenamefont {Freund},
  \citenamefont {Jesche}, \citenamefont {Gegenwart}, \citenamefont {Eschmann},
  \citenamefont {Buessen}, \citenamefont {Trebst}, \citenamefont
  {Van~Loosdrecht}, \citenamefont {Van Den~Brink},\ and\ \citenamefont
  {Gr{\"{u}}ninger}}]{Revelli2020FingerprintsIridates}%
  \BibitemOpen
  \bibfield  {author} {\bibinfo {author} {\bibfnamefont {A.}~\bibnamefont
  {Revelli}}, \bibinfo {author} {\bibfnamefont {M.}~\bibnamefont
  {Moretti~Sala}}, \bibinfo {author} {\bibfnamefont {G.}~\bibnamefont
  {Monaco}}, \bibinfo {author} {\bibfnamefont {C.}~\bibnamefont {Hickey}},
  \bibinfo {author} {\bibfnamefont {P.}~\bibnamefont {Becker}}, \bibinfo
  {author} {\bibfnamefont {F.}~\bibnamefont {Freund}}, \bibinfo {author}
  {\bibfnamefont {A.}~\bibnamefont {Jesche}}, \bibinfo {author} {\bibfnamefont
  {P.}~\bibnamefont {Gegenwart}}, \bibinfo {author} {\bibfnamefont
  {T.}~\bibnamefont {Eschmann}}, \bibinfo {author} {\bibfnamefont {F.~L.}\
  \bibnamefont {Buessen}}, \bibinfo {author} {\bibfnamefont {S.}~\bibnamefont
  {Trebst}}, \bibinfo {author} {\bibfnamefont {P.~H.~M.}\ \bibnamefont
  {Van~Loosdrecht}}, \bibinfo {author} {\bibfnamefont {J.}~\bibnamefont {Van
  Den~Brink}},\ and\ \bibinfo {author} {\bibfnamefont {M.}~\bibnamefont
  {Gr{\"{u}}ninger}},\ }\bibfield  {title} {\bibinfo {title} {{Fingerprints of
  Kitaev physics in the magnetic excitations of honeycomb iridates}},\ }\href
  {https://doi.org/10.1103/PhysRevResearch.2.043094} {\bibfield  {journal}
  {\bibinfo  {journal} {Physical Review Research}\ }\textbf {\bibinfo {volume}
  {2}},\ \bibinfo {pages} {43094} (\bibinfo {year} {2020})}\BibitemShut
  {NoStop}%
\bibitem [{\citenamefont {Revelli}\ \emph {et~al.}(2022)\citenamefont
  {Revelli}, \citenamefont {Moretti~Sala}, \citenamefont {Monaco},
  \citenamefont {Magnaterra}, \citenamefont {Attig}, \citenamefont {Peterlini},
  \citenamefont {Dey}, \citenamefont {Tsirlin}, \citenamefont {Gegenwart},
  \citenamefont {Fr\"ohlich}, \citenamefont {Braden}, \citenamefont {Grams},
  \citenamefont {Hemberger}, \citenamefont {Becker}, \citenamefont {van
  Loosdrecht}, \citenamefont {Khomskii}, \citenamefont {van~den Brink},
  \citenamefont {Hermanns},\ and\ \citenamefont
  {Gr\"uninger}}]{Revelli2022Quasimolecular}%
  \BibitemOpen
  \bibfield  {author} {\bibinfo {author} {\bibfnamefont {A.}~\bibnamefont
  {Revelli}}, \bibinfo {author} {\bibfnamefont {M.}~\bibnamefont
  {Moretti~Sala}}, \bibinfo {author} {\bibfnamefont {G.}~\bibnamefont
  {Monaco}}, \bibinfo {author} {\bibfnamefont {M.}~\bibnamefont {Magnaterra}},
  \bibinfo {author} {\bibfnamefont {J.}~\bibnamefont {Attig}}, \bibinfo
  {author} {\bibfnamefont {L.}~\bibnamefont {Peterlini}}, \bibinfo {author}
  {\bibfnamefont {T.}~\bibnamefont {Dey}}, \bibinfo {author} {\bibfnamefont
  {A.~A.}\ \bibnamefont {Tsirlin}}, \bibinfo {author} {\bibfnamefont
  {P.}~\bibnamefont {Gegenwart}}, \bibinfo {author} {\bibfnamefont
  {T.}~\bibnamefont {Fr\"ohlich}}, \bibinfo {author} {\bibfnamefont
  {M.}~\bibnamefont {Braden}}, \bibinfo {author} {\bibfnamefont
  {C.}~\bibnamefont {Grams}}, \bibinfo {author} {\bibfnamefont
  {J.}~\bibnamefont {Hemberger}}, \bibinfo {author} {\bibfnamefont
  {P.}~\bibnamefont {Becker}}, \bibinfo {author} {\bibfnamefont {P.~H.~M.}\
  \bibnamefont {van Loosdrecht}}, \bibinfo {author} {\bibfnamefont {D.~I.}\
  \bibnamefont {Khomskii}}, \bibinfo {author} {\bibfnamefont {J.}~\bibnamefont
  {van~den Brink}}, \bibinfo {author} {\bibfnamefont {M.}~\bibnamefont
  {Hermanns}},\ and\ \bibinfo {author} {\bibfnamefont {M.}~\bibnamefont
  {Gr\"uninger}},\ }\bibfield  {title} {\bibinfo {title} {Quasimolecular
  electronic structure of the spin-liquid candidate
  {Ba}$_{3}${InIr}$_{2}${O}$_{9}$},\ }\href
  {https://doi.org/10.1103/PhysRevB.106.155107} {\bibfield  {journal} {\bibinfo
   {journal} {Phys. Rev. B}\ }\textbf {\bibinfo {volume} {106}},\ \bibinfo
  {pages} {155107} (\bibinfo {year} {2022})}\BibitemShut {NoStop}%
\bibitem [{\citenamefont {Silinskas}\ \emph {et~al.}(2024)\citenamefont
  {Silinskas}, \citenamefont {Senz}, \citenamefont {Gargiani}, \citenamefont
  {Kalkofen}, \citenamefont {Kostanovskiy}, \citenamefont {Mohseni},
  \citenamefont {Meyerheim}, \citenamefont {Parkin},\ and\ \citenamefont
  {Bedoya-Pinto}}]{SilinskasOnFilms}%
  \BibitemOpen
  \bibfield  {author} {\bibinfo {author} {\bibfnamefont {M.}~\bibnamefont
  {Silinskas}}, \bibinfo {author} {\bibfnamefont {S.}~\bibnamefont {Senz}},
  \bibinfo {author} {\bibfnamefont {P.}~\bibnamefont {Gargiani}}, \bibinfo
  {author} {\bibfnamefont {B.}~\bibnamefont {Kalkofen}}, \bibinfo {author}
  {\bibfnamefont {I.}~\bibnamefont {Kostanovskiy}}, \bibinfo {author}
  {\bibfnamefont {K.}~\bibnamefont {Mohseni}}, \bibinfo {author} {\bibfnamefont
  {H.~L.}\ \bibnamefont {Meyerheim}}, \bibinfo {author} {\bibfnamefont
  {S.~S.~P.}\ \bibnamefont {Parkin}},\ and\ \bibinfo {author} {\bibfnamefont
  {A.}~\bibnamefont {Bedoya-Pinto}},\ }\href {https://arxiv.org/abs/2309.17439}
  {\bibinfo {title} {Self-intercalation as origin of high-temperature
  ferromagnetism in epitaxially grown {Fe}$_5${GeTe}$_2$ thin films}} (\bibinfo
  {year} {2024}),\ \Eprint {https://arxiv.org/abs/2309.17439} {arXiv:2309.17439
  [cond-mat.mtrl-sci]} \BibitemShut {NoStop}%
\bibitem [{\citenamefont {Gao}\ \emph {et~al.}(2020)\citenamefont {Gao},
  \citenamefont {Yin}, \citenamefont {Wang}, \citenamefont {Li}, \citenamefont
  {Cai}, \citenamefont {Zhao}, \citenamefont {Lei}, \citenamefont {Wang},
  \citenamefont {Zhang},\ and\ \citenamefont
  {Shen}}]{Gao2020SpontaneousFe5xGeTe2}%
  \BibitemOpen
  \bibfield  {author} {\bibinfo {author} {\bibfnamefont {Y.}~\bibnamefont
  {Gao}}, \bibinfo {author} {\bibfnamefont {Q.}~\bibnamefont {Yin}}, \bibinfo
  {author} {\bibfnamefont {Q.}~\bibnamefont {Wang}}, \bibinfo {author}
  {\bibfnamefont {Z.}~\bibnamefont {Li}}, \bibinfo {author} {\bibfnamefont
  {J.}~\bibnamefont {Cai}}, \bibinfo {author} {\bibfnamefont {T.}~\bibnamefont
  {Zhao}}, \bibinfo {author} {\bibfnamefont {H.}~\bibnamefont {Lei}}, \bibinfo
  {author} {\bibfnamefont {S.}~\bibnamefont {Wang}}, \bibinfo {author}
  {\bibfnamefont {Y.}~\bibnamefont {Zhang}},\ and\ \bibinfo {author}
  {\bibfnamefont {B.}~\bibnamefont {Shen}},\ }\bibfield  {title} {\bibinfo
  {title} {Spontaneous (anti)meron chains in the domain walls of van der waals
  ferromagnetic {Fe}$_{5-x}${GeTe}$_2$},\ }\href
  {https://doi.org/https://doi.org/10.1002/adma.202005228} {\bibfield
  {journal} {\bibinfo  {journal} {Advanced Materials}\ }\textbf {\bibinfo
  {volume} {32}},\ \bibinfo {pages} {2005228} (\bibinfo {year}
  {2020})}\BibitemShut {NoStop}%
\bibitem [{\citenamefont {Wu}\ \emph {et~al.}(2023)\citenamefont {Wu},
  \citenamefont {Basak}, \citenamefont {Li}, \citenamefont {Kim}, \citenamefont
  {Ryan}, \citenamefont {Lu}, \citenamefont {Hashimoto}, \citenamefont
  {Nelson}, \citenamefont {Acevedo-Esteves}, \citenamefont {Haley},
  \citenamefont {Analytis}, \citenamefont {He}, \citenamefont {Frano},\ and\
  \citenamefont {Birgeneau}}]{Shan2023chargeorder}%
  \BibitemOpen
  \bibfield  {author} {\bibinfo {author} {\bibfnamefont {S.}~\bibnamefont
  {Wu}}, \bibinfo {author} {\bibfnamefont {R.}~\bibnamefont {Basak}}, \bibinfo
  {author} {\bibfnamefont {W.}~\bibnamefont {Li}}, \bibinfo {author}
  {\bibfnamefont {J.-W.}\ \bibnamefont {Kim}}, \bibinfo {author} {\bibfnamefont
  {P.~J.}\ \bibnamefont {Ryan}}, \bibinfo {author} {\bibfnamefont
  {D.}~\bibnamefont {Lu}}, \bibinfo {author} {\bibfnamefont {M.}~\bibnamefont
  {Hashimoto}}, \bibinfo {author} {\bibfnamefont {C.}~\bibnamefont {Nelson}},
  \bibinfo {author} {\bibfnamefont {R.}~\bibnamefont {Acevedo-Esteves}},
  \bibinfo {author} {\bibfnamefont {S.~C.}\ \bibnamefont {Haley}}, \bibinfo
  {author} {\bibfnamefont {J.~G.}\ \bibnamefont {Analytis}}, \bibinfo {author}
  {\bibfnamefont {Y.}~\bibnamefont {He}}, \bibinfo {author} {\bibfnamefont
  {A.}~\bibnamefont {Frano}},\ and\ \bibinfo {author} {\bibfnamefont {R.~J.}\
  \bibnamefont {Birgeneau}},\ }\bibfield  {title} {\bibinfo {title} {Discovery
  of charge order in the transition metal dichalcogenide
  {Fe}$_{x}${NbS}$_{2}$},\ }\href
  {https://doi.org/10.1103/PhysRevLett.131.186701} {\bibfield  {journal}
  {\bibinfo  {journal} {Phys. Rev. Lett.}\ }\textbf {\bibinfo {volume} {131}},\
  \bibinfo {pages} {186701} (\bibinfo {year} {2023})}\BibitemShut {NoStop}%
\bibitem [{\citenamefont {Brown}\ \emph {et~al.}(2005)\citenamefont {Brown},
  \citenamefont {Fradkin},\ and\ \citenamefont
  {Kivelson}}]{Brown2005SurfaceTransition}%
  \BibitemOpen
  \bibfield  {author} {\bibinfo {author} {\bibfnamefont {S.~E.}\ \bibnamefont
  {Brown}}, \bibinfo {author} {\bibfnamefont {E.}~\bibnamefont {Fradkin}},\
  and\ \bibinfo {author} {\bibfnamefont {S.~A.}\ \bibnamefont {Kivelson}},\
  }\bibfield  {title} {\bibinfo {title} {Surface pinning of fluctuating charge
  order: An extraordinary surface phase transition},\ }\href
  {https://doi.org/10.1103/PhysRevB.71.224512} {\bibfield  {journal} {\bibinfo
  {journal} {Phys. Rev. B}\ }\textbf {\bibinfo {volume} {71}},\ \bibinfo
  {pages} {224512} (\bibinfo {year} {2005})}\BibitemShut {NoStop}%
\bibitem [{\citenamefont {Gu}\ \emph {et~al.}(2023)\citenamefont {Gu},
  \citenamefont {Carroll}, \citenamefont {Wang}, \citenamefont {Ran},
  \citenamefont {Broyles}, \citenamefont {Siddiquee}, \citenamefont {Butch},
  \citenamefont {Saha}, \citenamefont {Paglione}, \citenamefont {Davis},\ and\
  \citenamefont {Liu}}]{Gu2023DetectionUTe2}%
  \BibitemOpen
  \bibfield  {author} {\bibinfo {author} {\bibfnamefont {Q.}~\bibnamefont
  {Gu}}, \bibinfo {author} {\bibfnamefont {J.~P.}\ \bibnamefont {Carroll}},
  \bibinfo {author} {\bibfnamefont {S.}~\bibnamefont {Wang}}, \bibinfo {author}
  {\bibfnamefont {S.}~\bibnamefont {Ran}}, \bibinfo {author} {\bibfnamefont
  {C.}~\bibnamefont {Broyles}}, \bibinfo {author} {\bibfnamefont
  {H.}~\bibnamefont {Siddiquee}}, \bibinfo {author} {\bibfnamefont {N.~P.}\
  \bibnamefont {Butch}}, \bibinfo {author} {\bibfnamefont {S.~R.}\ \bibnamefont
  {Saha}}, \bibinfo {author} {\bibfnamefont {J.}~\bibnamefont {Paglione}},
  \bibinfo {author} {\bibfnamefont {J.~C.~S.}\ \bibnamefont {Davis}},\ and\
  \bibinfo {author} {\bibfnamefont {X.}~\bibnamefont {Liu}},\ }\bibfield
  {title} {\bibinfo {title} {{Detection of a pair density wave state in
  {UTe}$_2$}},\ }\href {https://www.nature.com/articles/s41586-023-05919-7}
  {\bibfield  {journal} {\bibinfo  {journal} {Nature}\ }\textbf {\bibinfo
  {volume} {618}},\ \bibinfo {pages} {921} (\bibinfo {year}
  {2023})}\BibitemShut {NoStop}%
\bibitem [{\citenamefont {Kengle}\ \emph {et~al.}(2024)\citenamefont {Kengle},
  \citenamefont {Vonka}, \citenamefont {Francoual}, \citenamefont {Chang},
  \citenamefont {Abbamonte}, \citenamefont {Janoschek}, \citenamefont {Rosa},\
  and\ \citenamefont {Simeth}}]{Kengle2024Absence2}%
  \BibitemOpen
  \bibfield  {author} {\bibinfo {author} {\bibfnamefont {C.~S.}\ \bibnamefont
  {Kengle}}, \bibinfo {author} {\bibfnamefont {J.}~\bibnamefont {Vonka}},
  \bibinfo {author} {\bibfnamefont {S.}~\bibnamefont {Francoual}}, \bibinfo
  {author} {\bibfnamefont {J.}~\bibnamefont {Chang}}, \bibinfo {author}
  {\bibfnamefont {P.}~\bibnamefont {Abbamonte}}, \bibinfo {author}
  {\bibfnamefont {M.}~\bibnamefont {Janoschek}}, \bibinfo {author}
  {\bibfnamefont {P.~F.~S.}\ \bibnamefont {Rosa}},\ and\ \bibinfo {author}
  {\bibfnamefont {W.}~\bibnamefont {Simeth}},\ }\href
  {https://arxiv.org/abs/2406.14690} {\bibinfo {title} {Absence of bulk charge
  density wave order in the normal state of {UTe}$_2$}} (\bibinfo {year}
  {2024}),\ \Eprint {https://arxiv.org/abs/2406.14690} {arXiv:2406.14690
  [cond-mat.str-el]} \BibitemShut {NoStop}%
\bibitem [{\citenamefont {Theuss}\ \emph {et~al.}(2024)\citenamefont {Theuss},
  \citenamefont {Shragai}, \citenamefont {Grissonnanche}, \citenamefont
  {Peralta}, \citenamefont {de~la Fuente~Simarro}, \citenamefont {Hayes},
  \citenamefont {Saha}, \citenamefont {Eo}, \citenamefont {Suarez},
  \citenamefont {Salinas}, \citenamefont {Pokharel}, \citenamefont {Wilson},
  \citenamefont {Butch}, \citenamefont {Paglione},\ and\ \citenamefont
  {Ramshaw}}]{Theuss2024Absence2}%
  \BibitemOpen
  \bibfield  {author} {\bibinfo {author} {\bibfnamefont {F.}~\bibnamefont
  {Theuss}}, \bibinfo {author} {\bibfnamefont {A.}~\bibnamefont {Shragai}},
  \bibinfo {author} {\bibfnamefont {G.}~\bibnamefont {Grissonnanche}}, \bibinfo
  {author} {\bibfnamefont {L.}~\bibnamefont {Peralta}}, \bibinfo {author}
  {\bibfnamefont {G.}~\bibnamefont {de~la Fuente~Simarro}}, \bibinfo {author}
  {\bibfnamefont {I.~M.}\ \bibnamefont {Hayes}}, \bibinfo {author}
  {\bibfnamefont {S.~R.}\ \bibnamefont {Saha}}, \bibinfo {author}
  {\bibfnamefont {Y.~S.}\ \bibnamefont {Eo}}, \bibinfo {author} {\bibfnamefont
  {A.}~\bibnamefont {Suarez}}, \bibinfo {author} {\bibfnamefont {A.~C.}\
  \bibnamefont {Salinas}}, \bibinfo {author} {\bibfnamefont {G.}~\bibnamefont
  {Pokharel}}, \bibinfo {author} {\bibfnamefont {S.~D.}\ \bibnamefont
  {Wilson}}, \bibinfo {author} {\bibfnamefont {N.~P.}\ \bibnamefont {Butch}},
  \bibinfo {author} {\bibfnamefont {J.}~\bibnamefont {Paglione}},\ and\
  \bibinfo {author} {\bibfnamefont {B.~J.}\ \bibnamefont {Ramshaw}},\ }\href
  {https://arxiv.org/abs/2406.14714} {\bibinfo {title} {Absence of a bulk
  thermodynamic phase transition to a density wave phase in {UTe}$_2$}}
  (\bibinfo {year} {2024}),\ \Eprint {https://arxiv.org/abs/2406.14714}
  {arXiv:2406.14714 [cond-mat.str-el]} \BibitemShut {NoStop}%
\bibitem [{\citenamefont {Mayer}\ \emph {et~al.}(2007)\citenamefont {Mayer},
  \citenamefont {Giannuzzi}, \citenamefont {Kamino},\ and\ \citenamefont
  {Michael}}]{Mayer2007TEMDamage}%
  \BibitemOpen
  \bibfield  {author} {\bibinfo {author} {\bibfnamefont {J.}~\bibnamefont
  {Mayer}}, \bibinfo {author} {\bibfnamefont {L.~A.}\ \bibnamefont
  {Giannuzzi}}, \bibinfo {author} {\bibfnamefont {T.}~\bibnamefont {Kamino}},\
  and\ \bibinfo {author} {\bibfnamefont {J.}~\bibnamefont {Michael}},\
  }\bibfield  {title} {\bibinfo {title} {{TEM Sample Preparation and
  FIB-Induced Damage}},\ }\href
  {https://link.springer.com/article/10.1557/mrs2007.63} {\bibfield  {journal}
  {\bibinfo  {journal} {MRS BULLETIN}\ }\textbf {\bibinfo {volume} {32}},\
  \bibinfo {pages} {401} (\bibinfo {year} {2007})}\BibitemShut {NoStop}%
\bibitem [{\citenamefont {Volkert}\ and\ \citenamefont
  {Minor}(2007)}]{Volkert2007FocusedMicromachining}%
  \BibitemOpen
  \bibfield  {author} {\bibinfo {author} {\bibfnamefont {C.~A.}\ \bibnamefont
  {Volkert}}\ and\ \bibinfo {author} {\bibfnamefont {A.~M.}\ \bibnamefont
  {Minor}},\ }\bibfield  {title} {\bibinfo {title} {Focused ion beam microscopy
  and micromachining},\ }\href {https://doi.org/10.1557/mrs2007.62} {\bibfield
  {journal} {\bibinfo  {journal} {MRS Bulletin}\ }\textbf {\bibinfo {volume}
  {32}},\ \bibinfo {pages} {389} (\bibinfo {year} {2007})}\BibitemShut
  {NoStop}%
\bibitem [{\citenamefont {Dvorak}\ \emph {et~al.}(2016)\citenamefont {Dvorak},
  \citenamefont {Jarrige}, \citenamefont {Bisogni}, \citenamefont {Coburn},\
  and\ \citenamefont {Leonhardt}}]{DovrakRSI2016}%
  \BibitemOpen
  \bibfield  {author} {\bibinfo {author} {\bibfnamefont {J.}~\bibnamefont
  {Dvorak}}, \bibinfo {author} {\bibfnamefont {I.}~\bibnamefont {Jarrige}},
  \bibinfo {author} {\bibfnamefont {V.}~\bibnamefont {Bisogni}}, \bibinfo
  {author} {\bibfnamefont {S.}~\bibnamefont {Coburn}},\ and\ \bibinfo {author}
  {\bibfnamefont {W.}~\bibnamefont {Leonhardt}},\ }\bibfield  {title} {\bibinfo
  {title} {{Towards 10 meV resolution: The design of an ultrahigh resolution
  soft X-ray RIXS spectrometer}},\ }\href
  {https://pubs.aip.org/aip/rsi/article/87/11/115109/362984/Towards-10-meV-resolution-The-design-of-an}
  {\bibfield  {journal} {\bibinfo  {journal} {Review of Scientific
  Instruments}\ }\textbf {\bibinfo {volume} {87}} (\bibinfo {year}
  {2016})}\BibitemShut {NoStop}%
\bibitem [{\citenamefont {Momma}\ and\ \citenamefont {Izumi}(2008)}]{VESTA}%
  \BibitemOpen
  \bibfield  {author} {\bibinfo {author} {\bibfnamefont {K.}~\bibnamefont
  {Momma}}\ and\ \bibinfo {author} {\bibfnamefont {F.}~\bibnamefont {Izumi}},\
  }\bibfield  {title} {\bibinfo {title} {{{\it VESTA}: a three-dimensional
  visualization system for electronic and structural analysis}},\ }\href
  {https://doi.org/10.1107/S0021889808012016} {\bibfield  {journal} {\bibinfo
  {journal} {Journal of Applied Crystallography}\ }\textbf {\bibinfo {volume}
  {41}},\ \bibinfo {pages} {653} (\bibinfo {year} {2008})}\BibitemShut
  {NoStop}%
\end{thebibliography}%

\end{document}


\title{Supplementary information for ``Magnetic excitations and absence of charge order in the cleavable ferromagnet \FGT"}
\author{V. K. Bhartiya}
\email{vbhartiya1@bnl.gov}
\affiliation{\NSLSII}

\author{T. Kim}%
\affiliation{\NSLSII}
\author{J. Li}%
\affiliation{\NSLSII}

\author{T. P. Darlington}%
\affiliation{\ColumbiaUni}
\author{D. J. Rizzo}%
\affiliation{\ColumbiaUni}

\author{Y. Gu}%
\affiliation{\NSLSII}
\author{S. Fan}%
\affiliation{\NSLSII}

\author{C. Nelson}
\affiliation{\NSLSII}
\author{J. W. Freeland}
\affiliation{\APS}

\author{X. Xu}%
\affiliation{Department of Physics, University of Washington, Seatle, Washington 98195, USA.}
\author{D. N. Basov}%
\affiliation{\ColumbiaUni}
\author{J. Pelliciari}%
\affiliation{\NSLSII}
\author{A. F. May}%
\affiliation{Materials Science and Technology Division, Oak Ridge National Laboratory, Oak Ridge, Tennessee 37831, USA.}

\author{C. Mazzoli}
\affiliation{\NSLSII}
\author{V. Bisogni}%
\email{bisogni@bnl.gov}
\affiliation{\NSLSII}

\maketitle
\tableofcontents
\clearpage
\section{Phonons}
\begin{figure}[h]
  \includegraphics[width=0.4\textwidth]{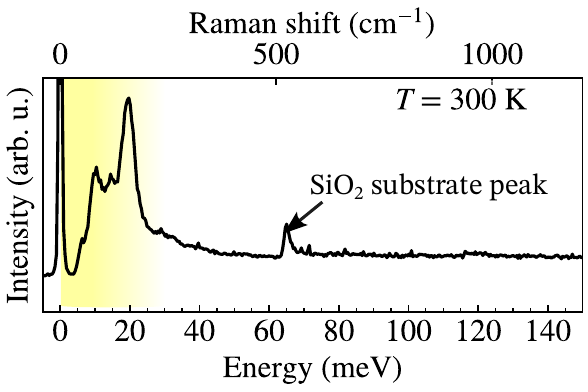}\\
  \caption{Raman scattering from an exfoliated flake. The  contribution from phonons to the spectrum is $\leq 20$ meV, highlighted by yellow color region.}\label{FIG:phonons}
\end{figure}

 The strongest phonon, at room temperature, is observed  below $<$ 150 cm$^{-1}$ (20 meV) i.e. no overlap with magnon or continuum (Supplementary Figure \ref{FIG:phonons}). We assume 20 meV phonon to have the dominant contribution in the \gls*{RIXS} spectrum. Raman lines around 62 meV originates from Si/SiO$_2$ substrate.

\section{Self-absorption correction to the RIXS intensity} \label{sec:abs_correction}

Outgoing photons with energy close to the absorption/resonance edge get considerably reabsorbed. Moreover, the path length traveled or the time spent by the photons inside the sample depends on the scattering geometry. Both effects combined affect the measured RIXS intensity, and when comparing absolute signal of two or more scattering geometries, it is essential to consider these aspects. The raw RIXS intensity can be corrected, to the first approximation, for these effects with information of X-ray absorption measured in total electron yield (TEY) and the corresponding scattering geometry \cite{Wang_Saturation_2020}. RIXS data shown in the main body of the paper was corrected for self-absorption effects following the procedure used in Refs.~\cite{Wang_Saturation_2020,Robarts2021DynamicalScattering}. Only scattering geometry lead self-absorption effects are considered as no dichroism is observed in TEY -- similar absorption profile for $\sigma$ and $\pi$- polarizations.
The corrected RIXS intensity $I_c$ is given as following:

\begin{equation}
   I_{c} = I_{m} \circledast \mathcal{A}\big(\theta,\Omega,\mu(\omega_{in}),\mu(\omega_{out})\big)
\end{equation}
where $  I_{m}$ is  measured RIXS intensity and the self-absorption correcting factor $\mathcal{A}$ is defined as, 
\begin{equation}
   \mathcal{A} = \frac{\mu(\omega_{in})\sin{(\Omega-\theta)} + \mu(\omega_{out}) \sin\theta}{\sin{(\Omega-\theta)}} 
\end{equation}
$\mathcal{A}$ is a function of scattering geometry ($\Omega$ is angle between incoming and outgoing photon and $\theta$ is photon incidence angle measured from the sample surface as shown in Fig. 1(a) of the main text, TEY, and, incoming and outgoing photon energy, 
$\mu(\omega_{in})$ is the total electron-yield (TEY) maximum (at Fe $L_3$- resonance) and $\mu(\omega_{out})$ is TEY value at the corresponding $\omega_{out}$. TEY was measured in the same ($\theta-\Omega$) scattering geometry as RIXS.

\section{RIXS spectrum fitting procedure}\label{sec:Fitting}

\begin{figure}[h]
  \includegraphics[width=0.45\textwidth]{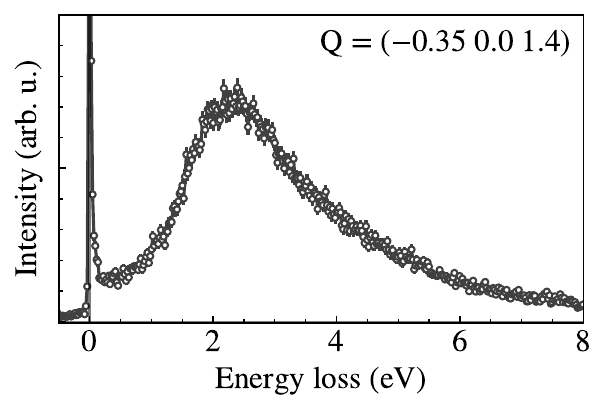}\\
  \caption{A representative raw RIXS spectrum. The fluorescence peak around 2 eV is a signature of metallic system. Its low energy part containing magnetic excitation (shown in Fig. 1(e) of the main text) is used for modeling.}\label{FIG:SingleRIXS}
\end{figure}

In this section we describe the fitting procedure of the self-absorption corrected low-energy portion [0 $-$ 300] meV RIXS intensity. A complete RIXS scan is shown in Supplementary Figure\ref{FIG:SingleRIXS}. \vb{RIXS spectra were normalized to the integral of the  fluorescence peak, before correcting for self-absorption.}  The elastic line was modeled by a Pseudo-Voigt model:

\begin{equation}
\begin{split}
    f_1(x;A,\mu,\sigma,\alpha) &= \frac{(1-\alpha)A}{\sigma_g \sqrt{2\pi}}e^{-(x-\mu)^2/2\sigma_g^2} \\&+ \frac{\alpha A}{\pi} \frac{\sigma}{(x-\mu)^2+\sigma^2}
    \end{split}
\end{equation}
 and the phonon and magnon by a Gaussian distribution:

\begin{equation}
    f_2(x;A,\mu,\sigma) = \frac{A}{\sigma \sqrt{2 \pi}} e^{[-(x-\mu)^2/2\sigma^2]}
\end{equation}

while to accommodate the damped and broad character of the continumm, it was modeled by a Skewed Gaussian:
\begin{equation}
    f_3(x;A,\mu,\sigma,\gamma) = \frac{A}{\sigma \sqrt{2 \pi}} e^{[-(x-\mu)^2/2\sigma^2]} \{1+\text{erf}[\frac{\gamma(x-\mu)}{\sigma\sqrt{2}}] \}
\end{equation}

The fluorescence tail background was modeled by a step function broadened by the twice experimental resolution:   

\begin{equation}
    f_4(x,A_0,A,\mu,\sigma) =  A[1+\text{erf}((x-\mu)/\sigma)]/2 
\end{equation}

Here $A$ is the amplitude, $\mu$ is the center, $\sigma$ is the standard-deviation, $\alpha$ is the fraction, Final model describing the entire RIXS spectrum for $E < 300$ meV. As a result, function capturing the low energy RIXS spectrum is as follows --

\begin{equation}
\begin{split}
     I &= \text{Elastic}(f_1) + \text{Phonon} (f_2)  + \text{Magnon} (f_2)  \\  &+ \text{Continuum} (f_3) +  \text{Background} (f_4) 
\end{split}
\end{equation}\\

During fitting of all the \gls*{RIXS} spectra vs $\bm{Q}$, the following procedure was adopted. Elastic line: $ \sigma, \alpha$ was kept fixed and only amplitude $A$ and $\mu$ were allowed to change. $\sigma$ was fixed by the experimental resolution.  Phonon:  $\sigma$ was fixed by experimental resolution and $\mu$ was fixed to 20 meV from the Raman measurements and amplitude $A$ was allowed to change as $\sin^2(\pi L)$ \cite{Lin2070StronglyAnomaly} for in-plane scans, however, it was fixed to \vbb{$L = 3.23$} scan out-of-plane dependence. Magnon: Both $A$  and $\mu$ were allowed to change, but $\sigma$ was fixed to \vbb{30\%} higher value than the resolution to account for continuum contamination \vbb{for in-plane scans}, and for the out-of-plane scans, both $A$  and $\mu$ were fixed to \vbb{$L = 3.23$}.  Continuum: Both its width $\sigma$ and $\gamma$ were fixed,  however, its amplitude $A$ and $\mu$ were allowed to change. \vbb{For simplicity in out-of-plane scans continuum width $\sigma$ was fixed to  $L = 3.23$ and $\mu$ from in-plane scan}. Background amplitude $A$ was  fixed to $\text{mean}$ (200 $<$ $I$ $<$ 450).

\section{Continuum intensity modulation with and without self-absorption correction}

A comparison of robust magnetic continuum intensity modulation between raw and self-absorption corrected RIXS spectrum is shown in Supplementary Figure  \ref{FIG:comparison}. 

\begin{figure}
  \includegraphics[width=0.4\textwidth]{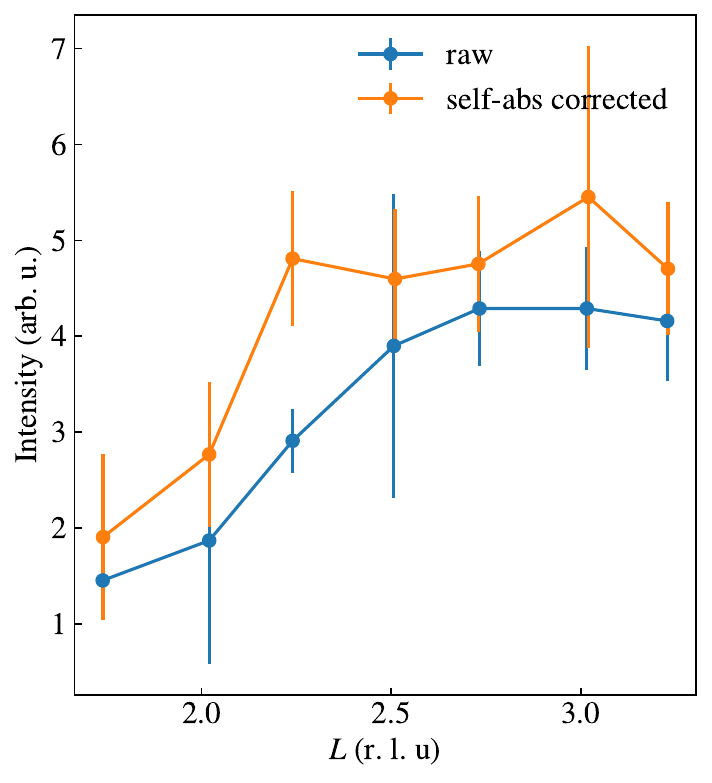}\\
  \caption{\vb{A comparison of the magnetic continuum intensity modulation. The scale of intensity variation is not significantly affected by the self-absorption correction. The raw RIXS spectra (blue) and self-absorption corrected spectra (orange) evolve in similar fashion, demonstrating $\approx$ 300 \% evolution. }} \label{FIG:comparison}
\end{figure}
\stoptoc
\bibliography{references_FGT}